\documentclass[11pt,prd,onecolumn,showpacs,amsmath,amssymb,aps,floats,floatfix,nofootinbib]{revtex4-1}

\usepackage{hyperref} 
\usepackage[inline]{enumitem}
\usepackage[multidot]{grffile}  
\usepackage{dcolumn}
\usepackage{bm}
\usepackage{amsmath}
\usepackage{amsfonts}
\usepackage{amssymb}
\usepackage{bbm}
\usepackage{color}
\usepackage{latexsym}
\usepackage{slashed} 
\usepackage{pstricks}
\usepackage{indentfirst}
\usepackage{mathrsfs}
\usepackage{multirow}
\usepackage{epsfig,psfrag}
\usepackage{subfigure}
\usepackage{setspace} 
\usepackage[utf8]{inputenc} 
\usepackage[scientific-notation=true]{siunitx} 

\usepackage{graphicx}
\usepackage{epstopdf}
\usepackage{epsfig}

\graphicspath{{figs/}}

\def\iab{\mathrm{ab}^{-1}} 
\def\TeV{\mathrm{TeV}}     
\def\GeV{\mathrm{GeV}}     
\def\MeV{\mathrm{MeV}}     
\def\keV{\mathrm{keV}}     
\def\missET{\slashed E_\mathrm{T}} 

\def\chia{{\chi}_1^0}
\def\chib{{\chi}_2^0}
\def\chic{{\chi}_3^0}

\def\chiaa{{\chi}_i^{0}}
\def\chii{{\chi}_i^{\pm}}
\def\chipm{{\chi}^{\pm}}
\def\chiapm{{\chi}_1^{\pm}}
\def\chibpm{{\chi}_2^{\pm}}
\def\mchia{m_{{\chi}_1^0}}
\def\mchib{m_{{\chi}_2^0}}

\def\mchipm{m_{{\chi}^{\pm}}}

\def\su2l{\mathrm{SU}(2)_\mathrm{L}}
\def\hc{\mathrm{h.c.}}
\def\mN{\mathcal{M}_\mathrm{N}}
\def\mC{\mathcal{M}_\mathrm{C}}
\def\N{\mathcal{N}}
\def\CL{\mathcal{C}_\mathrm{L}}
\def\CR{\mathcal{C}_\mathrm{R}}

\makeatother

\allowdisplaybreaks 

\begin{document}
\allowdisplaybreaks 
\setlength\arraycolsep{0.2em} 
\setstretch{1.2} 

\title{Exploring Fermionic Dark Matter via Higgs Boson Precision Measurements at the Circular Electron Positron Collider}
\author{Qian-Fei Xiang$^{1,2}$}
\author{Xiao-Jun Bi$^1$}
\author{Peng-Fei Yin$^1$}
\author{Zhao-Huan Yu$^{3,4}$}
\affiliation{$^1$Key Laboratory of Particle Astrophysics,
Institute of High Energy Physics, Chinese Academy of Sciences,
Beijing 100049, China}
\affiliation{$^2$School of Physical Sciences,
University of Chinese Academy of Sciences,
Beijing 100049, China}
\affiliation{$^3$ARC Centre of Excellence for Particle Physics at the Terascale,
School of Physics, The~University of Melbourne, Victoria 3010, Australia}
\affiliation{$^4$School of Physics, Sun Yat-Sen University, Guangzhou 510275, China}

\begin{abstract}
We study the impact of fermionic dark matter (DM) on projected Higgs precision measurements at the Circular Electron Positron Collider (CEPC), including the one-loop effects on the $e^+e^-\to Zh$ cross section and the Higgs boson diphoton decay, as well as the tree-level effects on the Higgs boson invisible decay.
As illuminating examples, we discuss two UV-complete DM models, whose dark sector contains electroweak multiplets that interact with the Higgs boson via Yukawa couplings.
The CEPC sensitivity to these models and current constraints from DM detection and collider experiments are investigated.
We find that there exist some parameter regions where the Higgs measurements at the CEPC will be complementary to current DM searches.
\end{abstract}
\pacs{12.15.Lk,12.60.Cn,13.66.Jn}
\maketitle
\tableofcontents

\clearpage

\section{Introduction}

The discovery of the Higgs boson at the Large Hadron Collider (LHC)~\cite{Aad:2012tfa,Chatrchyan:2012xdj} confirms the  particle content of the standard model (SM).
However, the existence of dark matter (DM)~\cite{Jungman:1995df,Bertone:2004pz,Feng:2010gw} undoubtedly implies the new physics beyond the SM (BSM).
While searches for new particles at the LHC will continue in the coming years, an alternative way to probe new physics is by studying its loop effects via high precision observables at $e^+e^-$ colliders.

Several electron-positron colliders have been currently proposed, including the Circular Electron Positron Collider (CEPC)~\cite{CEPC-SPPCStudyGroup:2015csa}, the Future Circular Collider with $e^+e^-$ collisions (FCC-ee)~\cite{Gomez-Ceballos:2013zzn}, and the International Linear Collider (ILC)~\cite{Baer:2013cma}.
These machines are planned to serve as ``Higgs factories'' for precisely measuring the properties of the Higgs boson.
In particular, CEPC will run at a center-of-mass energy of $240-250~\si{GeV}$, which maximizes the $e^+e^- \to Zh$ production, over ten years to collect a data set of $5~\iab$.

Exploiting the physics potential of the CEPC has attracted many interests.
Recent works for probing anomalous couplings include studies on the anomalous $hhh$ and $htt$ couplings through the $e^+ e^- \to Zh$ measurement~\cite{McCullough:2013rea,Shen:2015pha,Huang:2015izx, Kobakhidze:2016mfx}, the anomalous $hZ\gamma$ and $h\gamma\gamma$ couplings through the $e^+ e^- \to h \gamma$ measurement~\cite{Cao:2015iua,Hu:2014eia}, and the anomalous $Zbb$ coupling~\cite{Gori:2015nqa}, and high order effective operators~\cite{Fedderke:2015txa,Ge:2016zro}.
Other CEPC researches about new physics models involve studies on natural supersymmetry~\cite{Fan:2014axa,Cao:2016uwt,Wu:2017kgr}, DM models~\cite{Yu:2013aca,Yu:2014ula,Harigaya:2015yaa,Cao:2016qgc,Xiang:2016jni} and electroweak oblique parameters~\cite{Fedderke:2015txa,Cai:2016sjz,Cai:2017wdu}, and so on~\cite{Englert:2013tya,Cao:2014ita}.

In this work, we mainly study the impact of fermionic DM on the Higgs physics at the CEPC. Particularly, we focus on the loop effects on the $e^+e^- \to Zh$ production cross section, whose relative precision will be pinned down to $0.5\%$~\cite{CEPC-SPPCStudyGroup:2015csa}.
For this purpose, the DM particle should couple to both the Higgs and $Z$ bosons and modify the $hZZ$ coupling at one-loop level.
This requirement can be fulfilled by introducing
a dark sector consisting of electroweak multiplets, which is a simple, UV-complete extension to the SM.
Such a dark sector would provide an attractive DM candidate that naturally satisfies the observed relic abundance.
Related model buildings typically involve one $\su2l$ multiplet, which leads to the so-called minimal DM models~\cite{Cirelli:2005uq, Cirelli:2007xd, Cirelli:2009uv, Hambye:2009pw, Cai:2012kt, Ostdiek:2015aga, Cai:2015kpa},
or more than one $\su2l$ multiplet~\cite{Mahbubani:2005pt, DEramo:2007anh, Enberg:2007rp, Cohen:2011ec, Fischer:2013hwa, Cheung:2013dua, Dedes:2014hga, Fedderke:2015txa, Calibbi:2015nha, Freitas:2015hsa, Yaguna:2015mva, Tait:2016qbg, Horiuchi:2016tqw, Banerjee:2016hsk, Cai:2016sjz, Abe:2017glm, Cai:2017wdu, Arcadi:2017kky, Maru:2017otg, Liu:2017gfg, Egana-Ugrinovic:2017jib}.
As we would like to discuss fermionic DM, more than one multiplet is needed for allowing renormalizable couplings to the Higgs boson with respect to the gauge invariance.

We calculate one-loop corrections to $e^+e^- \to Zh$ contributed by the dark sector.
For the purpose of illustration, we study two simple models with additional fermionic $\su2l$ multiplets:
\begin{itemize}
	\item Singlet-doublet fermionic dark matter (SDFDM) model: the dark sector involves one singlet Weyl spinor and two doublet Weyl spinors;
	\item Doublet-triplet fermionic dark matter (DTFDM) model: the dark sector involves two doublet Weyl spinor and one triplet Weyl spinors.
\end{itemize}
These spinors are assumed to be vectorlike, in order to cancel gauge anomalies. This means that the two doublets should have opposite hypercharges, while the singlet or the triplet should have zero hypercharge.

\begin{table}[!tbp]
\centering
\setlength{\tabcolsep}{.4em}
\renewcommand{\arraystretch}{1.3}
\begin{tabular}{ccc}
\hline\hline
Models & Gauge eigenstates & Mass eigenstates\\
\hline
Singlet-Doublet
  & $S$,
     $\begin{pmatrix}
			  D_1^0 \\
			  D_1^-
			\end{pmatrix}$,
     $\begin{pmatrix}
			  D_2^+ \\
			  D_2^0
			\end{pmatrix}$
		& $\begin{matrix}
                {} \\[-1.2em]
                \chi_1^0, \chi_2^0, \chi_3^0 \\
                \chi^\pm \\[-1.2em]
                {}
            \end{matrix}$
\\
\hline
Doublet-Triplet
  & $\begin{pmatrix}
			  D_1^0 \\
			  D_1^-
        \end{pmatrix}$,
     $\begin{pmatrix}
			  D_2^+ \\
			  D_2^0
	   \end{pmatrix}$,
      $\begin{pmatrix}
			     T^+ \\
			     T^0 \\
			     -T^-
		 \end{pmatrix}$
  &   $\begin{matrix}
            {} \\[-.5em]
            \chi_1^0, \chi_2^0, \chi_3^0 \\
            \chi_1^\pm, \chi_2^\pm \\[-.5em]
            {}
    	 \end{matrix}$ \\
\hline\hline
\end{tabular}
\caption{Field contents of the two DM models under consideration.}
\label{tab:model}
\end{table}

After electroweak symmetry-breaking (EWSB), the vacuum expectation value (VEV) of the Higgs doublet provides Dirac mass terms to the dark multiplets, leading to state mixings.
Field contents in the gauge and mass bases for the two models are denoted in Table~\ref{tab:model}.
The lightest neutral eigenstate ($\chia$) in the dark sector serves as a Majorana DM candidate.
For ensuring the stability of $\chia$, we need to impose a $Z_2$ symmetry, under which all SM particles are even and dark sector particles are odd.
These models can be regarded as the generalizations of some electroweak sectors in supersymmetric models.
For instance, the SDFDM model is similar to the bino-Higgsino sector, while the DTFDM model is similar to the Higgsino-wino sector.

Serving as a DM candidate, $\chia$ should be consistent with the observed DM relic abundance~\cite{Planck:2015xua}.
The $\chia$ couplings to the $Z$ and Higgs bosons could induce spin-dependent and spin-independent scatterings between nuclei and DM, respectively. They would be constrained by direct detection experiments~\cite{Tan:2016zwf,Fu:2016ega}.
Besides, there are bounds from colliders experiments, such as bounds from the invisible decay of the $Z$ boson~\cite{ALEPH:2005ab}, from searches for charged particles at the LEP, and from the monojet searches at the LHC~\cite{Aad:2015zva}.
Moreover, dark sector particles may affect the invisible and diphoton decays of the Higgs boson, which will be precisely determined by CEPC~\cite{CEPC-SPPCStudyGroup:2015csa}.
In this work, we investigate both the CEPC prospect and current experimental constraints for the two DM models.

The paper is outlined as follows.
In Sec.~\ref{sec:SDFDM} we give a brief description of the SDFDM model, identify the parameter regions that could be explored by Higgs measurements at the CEPC, and study current constraints from DM detection and collider experiments.
In Sec.~\ref{sec:DTFDM}, we repeat the calculations, but for the DTFDM model.
Sec.~\ref{sec:conclu} contains our conclusions and discussions.

\section{Singlet-Doublet Fermionic Dark Matter}
\label{sec:SDFDM}

\subsection{Model details}

In the SDFDM model~\cite{Mahbubani:2005pt, DEramo:2007anh, Enberg:2007rp, Cohen:2011ec, Cheung:2013dua, Calibbi:2015nha, Horiuchi:2016tqw, Banerjee:2016hsk, Cai:2016sjz, Abe:2017glm}, we introduce a dark sector with one Weyl singlet and two $\su2l$ Weyl doublets obeying the $(\su2l,\mathrm{U}(1)_\mathrm{Y})$ gauge transformations:
\begin{equation}
	S \in (\mathbf{1}, 0),\quad
    D_1 \equiv {D_1^0 \choose D_1^-} \in \left(\mathbf{2}, -\frac{1}{2}\right),\quad
    D_2 \equiv {D_2^+ \choose D_2^0} \in \left(\mathbf{2}, \frac{1}{2}\right).
\end{equation}
Here, the assignment of opposite hypercharges to the two doublets is essential to cancel the gauge anomalies.
We can write down the following gauge invariant Lagrangians:
\begin{eqnarray}
	\mathcal{L}_S &=& i S^\dagger \bar{\sigma}^\mu \partial_\mu S - \frac{1}{2} (m_S S S + \hc),\\
{{\cal L}_{\rm{D}}} &=& iD_1^\dag {{\bar \sigma }^\mu }{D_\mu }{D_1} + iD_2^\dag {{\bar \sigma }^\mu }{D_\mu }{D_2} - ({m_D} \epsilon_{ij} {D_1^i}{D_2^j} + \hc),
\end{eqnarray}
where $D_\mu =\partial_\mu - ig W_\mu^a \tau_a^{(2)} - i g' Y B_\mu$, with the generators $\tau_a^{(2)}=\sigma^a/2$ expressed by the Pauli matrices $\sigma^a$.
More specifically, gauge interactions of the doublets are given by
\begin{eqnarray}
  \mathcal{L} &\supset&
  \frac{{g}}{{2c_\mathrm{W}}} Z_\mu \left[ {(D{_1^0})^\dag {{\bar \sigma }^\mu }D_1^0 - (D{_2^0})^\dag {{\bar \sigma }^\mu }D_2^0 - \left( {1 - 2s_\mathrm{W}^2} \right)(D_1^-)^\dag {{\bar \sigma }^\mu }D_1^ -  + \left( {1 - 2s_\mathrm{W}^2} \right)(D_2^+)^\dag {{\bar \sigma }^\mu }D_2^ + } \right]	 \nonumber\\
  && +\frac{g}{{\sqrt 2 }}W_\mu^+ \left[ {(D{_1^0})^\dag {{\bar \sigma }^\mu }D_1^ -  + (D_2^+)^\dag {{\bar \sigma }^\mu }D_2^0} \right] + \frac{g}{{\sqrt 2 }}W_\mu^- \left[ {(D_1^-)^\dag {{\bar \sigma }^\mu }D_1^0 + (D{_2^0})^\dag {{\bar \sigma }^\mu }D_2^ + } \right] \nonumber\\
  && - e{A_\mu }\left[ {(D_1^-)^\dag {{\bar \sigma }^\mu }D_1^-  - (D_2^+)^\dag {{\bar \sigma }^\mu }D_2^ + } \right],
\label{eq:2gau}
\end{eqnarray}
where $c_\mathrm{W}\equiv\cos\theta_\mathrm{W}$ and $s_\mathrm{W}\equiv\sin\theta_\mathrm{W}$ are related to the Weinberg angle $\theta_\mathrm{W}$.
The dark sector fields interact with the SM Higgs doublet $H$ through the Yukawa couplings
\begin{equation}
 {{\cal L}_{{\rm{Y}}}} = {y_1}SD_1^i{H_i} - {y_2}SD_2^iH_i^\dag  + \hc
\end{equation}

After the EWSB, dark sector fermions obtain Dirac mass terms through the Higgs mechanism.
In the unitary gauge, $H = \big(0, (v + h)/\sqrt{2}\big)^\mathrm{T}$ with the VEV $v$.
The mass terms in the model can be expressed as
\begin{equation}
		\mathcal{L}_\mathrm{M} =
		 -\frac{1}{2}
		\left(
			\begin{array}{ccc}
				S & D_1^0 & D_2^0
			\end{array}
		\right)
		\mathcal{M}_\mathrm{N}
		\left(
			\begin{array}{c}
				S \\ D_1^0 \\ D_2^0
			\end{array}
		\right)
	- m_D D_1^- D_2^+ + \hc
	=  - \frac{1}{2} \sum_{i=1}^3 m_{\chiaa} \chiaa \chiaa - m_{\chi^\pm} \chi^- \chi^+ + \hc,
\end{equation}
where $\chi^- \equiv D_1^-$, $\chi^+ \equiv D_2^+$, and $m_{\chi^\pm} \equiv m_D$.
The mass matrix of the neutral states $\mathcal{M}_\mathrm{N}$ and the corresponding mixing matrix $\mathcal{N}$ to diagonalize it are given by
\begin{equation}
	\mN =
	\left(
		\begin{array}{ccc}
			m_S                      & \dfrac{1}{\sqrt{2}} y_1 v & \dfrac{1}{\sqrt{2}} y_2 v \\[1em]
			\dfrac{1}{\sqrt{2}} y_1 v & 0                        & -m_D \\[1em]
			 \dfrac{1}{\sqrt{2}} y_2 v & -m_D                    &0
		 \end{array}
	\right),\quad
	\N^\mathrm{T} \mN \N = \mathrm{diag}(m_{\chia}, m_{\chib}, m_{\chic}),\quad
	\left(
	\begin{array}{c}
		S \\ D_1^0 \\ D_2^0
	\end{array}
	\right)
	=
	\N
	\left(
	\begin{array}{c}
		\chia \\ \chib \\ \chic
	\end{array}
	\right).
\end{equation}
Thus, the dark sector contains one charged Dirac fermion $\chi^\pm$ and three Majorana fermions $\chi_{1,2,3}^0$, with the lightest neutral fermion $\chi_1^0$ serving as the DM particle.

This model is totally determined by four parameters, $y_1$, $y_2$, $m_S$, and $m_D$.
In principle, all of them could be complex and induce $CP$ violation. However, three phases can be eliminated by redefinition of the fields, leaving only one independent $CP$ violation phase.
The effects of this $CP$ violation phase on electric dipole moments and on DM direct detection have been studied by several groups~\cite{Mahbubani:2005pt,DEramo:2007anh,Abe:2017glm}.
We do not discuss these effects further, and take all parameters to be real below.

\begin{figure}[!tbp]
\centering
\subfigure[~$m_S = 100~\GeV$, $m_D = 400~\GeV$, $y_1=1$.]
{\includegraphics[width=.45\textwidth]{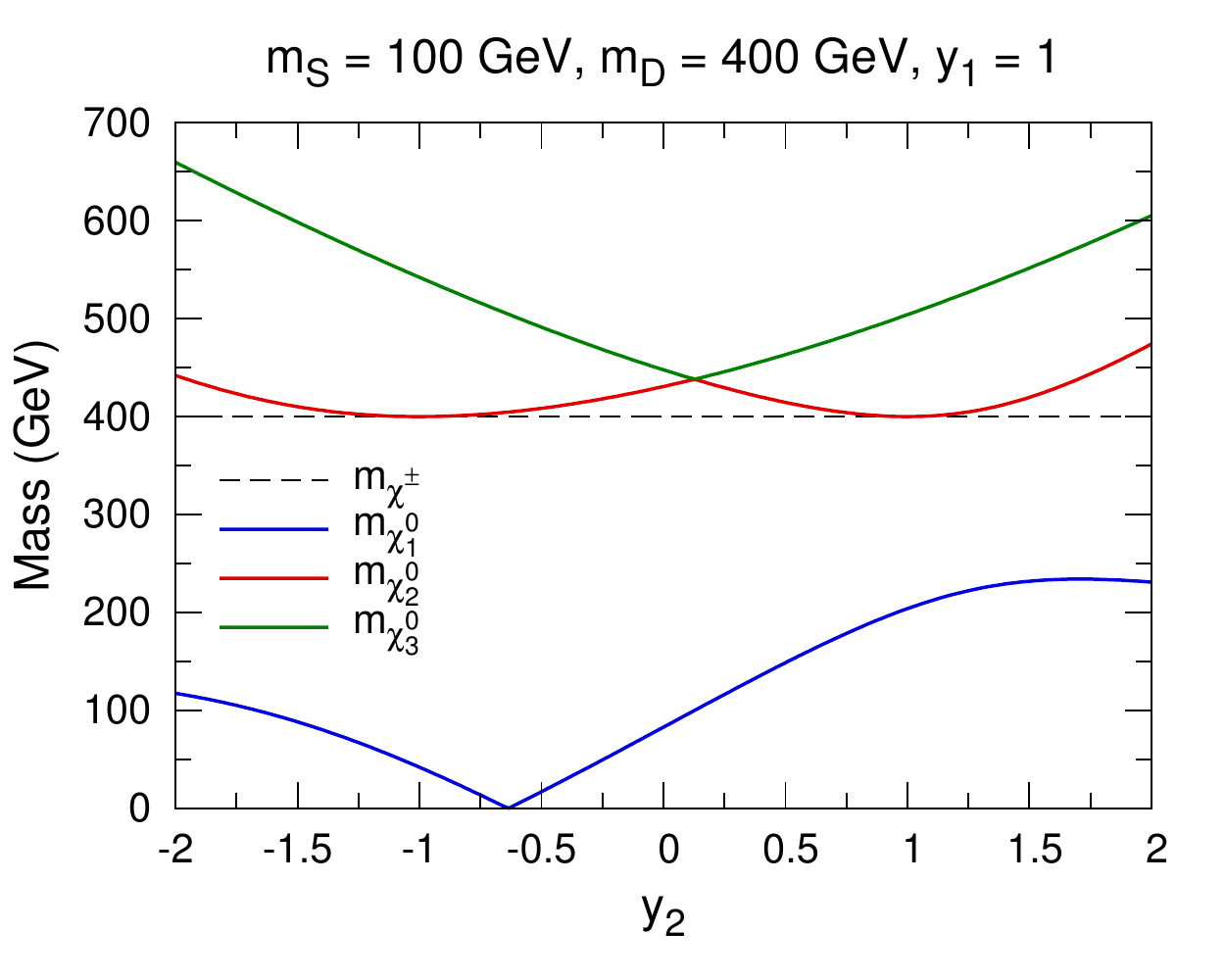}}
\subfigure[~$m_S = 400~\GeV$, $m_D = 150~\GeV$, $y_1=1$.]
{\includegraphics[width=.45\textwidth]{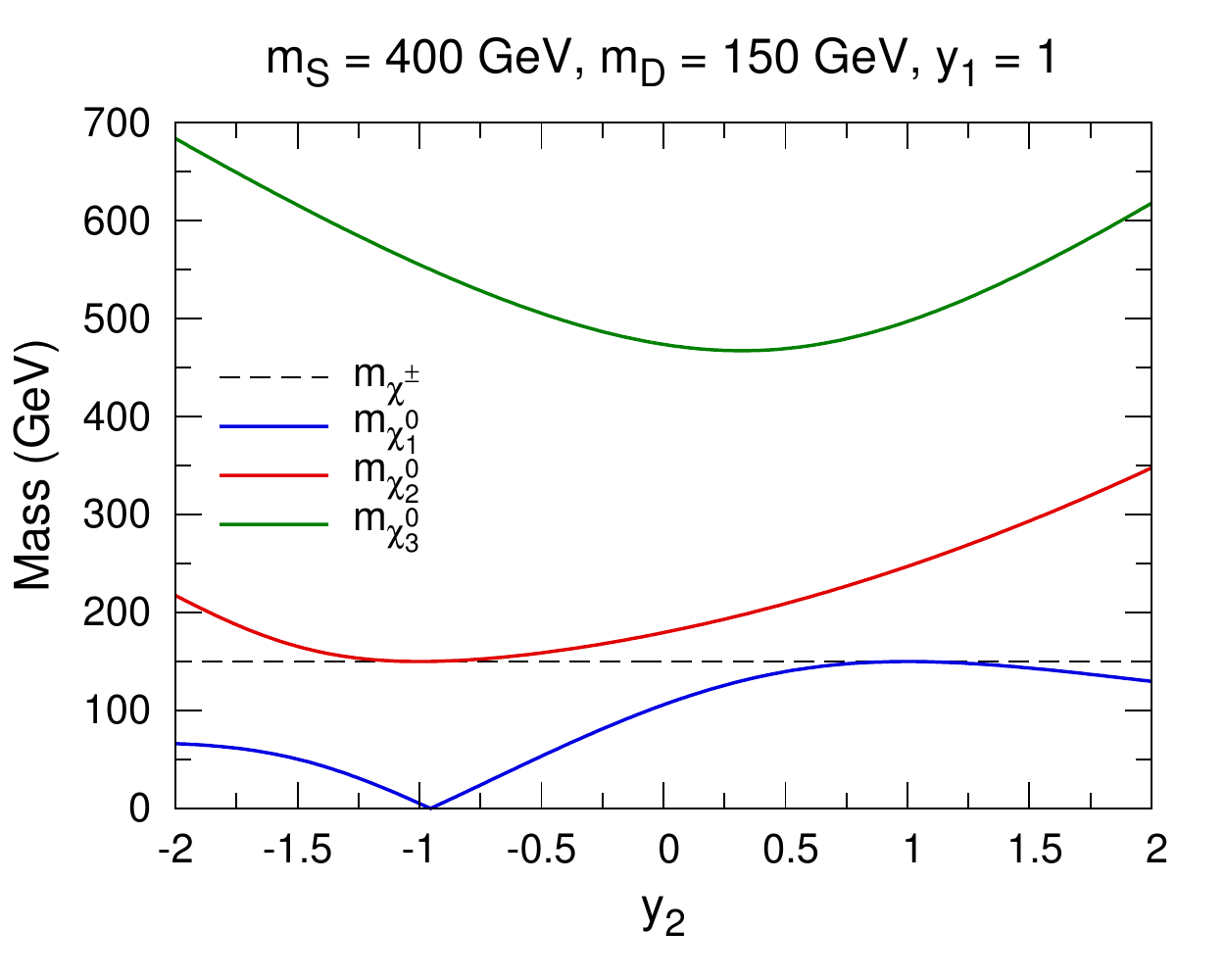}}
\caption{Mass spectra of the SDFDM model in two typical cases, $m_S < m_D$ (a) and $m_S > m_D$ (b).}
\label{fig:12spec}
\end{figure}

In Fig.~\ref{fig:12spec} we show the masses of the dark sector fermions as functions of $y_2$ with $y_1=1$ for two typical cases, $m_S < m_D$ and $m_S > m_D$.
If $m_S < m_D$, $\chia$ is singlet-dominated, with a mass close to $m_S$ when $y_1$ and $y_2$ are small;  $\chib$ and $\chic$ are doublet-dominated, with masses close to $m_D$ for small Yukawa couplings.
On the other hand, if $m_S > m_D$, $\chia$ and $\chib$ are doublet-dominated, while $\chic$ is singlet-dominated.
When $y_2=\pm y_1$, we have $\mchipm=\mchib$ or $\mchipm=\mchia$ due to a custodial symmetry.

It is instructive to reform the interaction terms with four-component spinors.
Defining Dirac spinor $\Psi^+$ and Majorana spinors $\Psi_i$ ($i=1,2,3$) as
\begin{equation}
\Psi^+ = \begin{pmatrix}
\chi^+\\
(\chi^-)^\dag
\end{pmatrix},\quad
\Psi_i = \begin{pmatrix}
\chi_i^0\\
(\chi_i^0)^\dag
\end{pmatrix},
\end{equation}
we have
\begin{eqnarray}
	\mathcal{L}_\mathrm{int}& = & e A_\mu \bar{\Psi}^+ \gamma^\mu \Psi^+
	+ \frac{g}{ 2 c_\mathrm{W}} (c_\mathrm{W}^2 - s_\mathrm{W}^2) Z_\mu \bar{\Psi}^+ \gamma^\mu \Psi^+ \nonumber\\
	&&+ \frac{g}{\sqrt{2}} \sum_i W_\mu^- (\N_{3i}^\ast \bar{\Psi}_i \gamma^\mu P_\mathrm{L} \Psi^+ - \N_{2i} \bar{\Psi}_i \gamma^\mu P_\mathrm{R} \Psi^+) \nonumber\\
	&&+ \frac{g}{\sqrt{2}} \sum_i W_\mu^+ (\N_{3i} \bar{\Psi}^+ \gamma^\mu P_\mathrm{L} \Psi_i - \N_{2i}^\ast \bar{\Psi}^+ \gamma^\mu P_\mathrm{R} \Psi_i) \nonumber\\
	&& -\frac{1}{2} \sum_{ij} C_{Z,ij}^\mathrm{A} Z_\mu \bar{\Psi}_i \gamma^\mu \gamma^5 \Psi_j
	  + \frac{1}{2} \sum_{ij} C_{Z,ij}^\mathrm{V} Z_\mu \bar{\Psi}_i  \gamma^\mu \Psi_j \nonumber\\
	&& -\frac{1}{2} \sum_{ij} C_{h,ij}^\mathrm{S} h \bar{\Psi}_i \Psi_j
	  +\frac{1}{2} \sum_{ij} C_{h,ij}^\mathrm{P} h \bar{\Psi}_i i \gamma^5 \Psi_j,
\end{eqnarray}
where $P_\mathrm{L}\equiv (1-\gamma^5)/2$ and $P_\mathrm{R}\equiv (1+\gamma^5)/2$.
The couplings to $Z$ and $h$ are given by
\begin{eqnarray}
	C_{Z,ij}^\mathrm{A} &=& \frac{g}{2 c_\mathrm{W}} \mathrm{Re}(\N_{2i}^\ast \N_{2j} - \N_{3i}^\ast \N_{3j}),\quad
	C_{Z,ij}^\mathrm{V} = \frac{ig}{2 c_\mathrm{W}} \mathrm{Im}(\N_{2i}^\ast \N_{2j} - \N_{3i}^\ast \N_{3j}),\\
	C_{h,ij}^\mathrm{S} &=& \sqrt{2}~\mathrm{Re}(y_1 \N_{1i} \N_{2j} + y_2 \N_{1i} \N_{3j}),\quad
	C_{h,ij}^\mathrm{P} = \sqrt{2}~\mathrm{Im}(y_1 \N_{1i} \N_{2j} + y_2 \N_{1i} \N_{3j}).
\end{eqnarray}

It is obvious to find that $C_{Z,ii}^\mathrm{V}=0$, due to the Majorana nature of $\Psi_i$.
Since $y_1$ and $y_2$ are real parameters, the $CP$-violating couplings $C_{h,ii}^\mathrm{P}$ also vanish.
For DM phenomenology, the $C_{Z,11}^\mathrm{A}$ and $C_{h,11}^\mathrm{S}$ couplings are particularly important, inducing spin-dependent (SD) and spin-independent (SI) DM-nucleon scattering, respectively. Therefore, they could be probed in direct detection experiments.

When $y_1 = \pm y_2$, there is a custodial global symmetry resulting $C_{Z,11}^\mathrm{A}=0$ and a vanishing SD scattering cross section.
Besides, if $m_D < m_S$, the condition $y_1 = y_2$ also leads to $C_{h,11}^\mathrm{S}=0$ and a vanishing SI cross section~\cite{Cai:2016sjz}.
It would be useful to explore other conditions that give rise to $C_{h,11}^\mathrm{S} =0$, which implies blind spots in direct detection experiments~\cite{Cohen:2011ec, Cheung:2012qy, Cheung:2013dua, Abe:2017glm}.
According to the low-energy Higgs theorems~\cite{Ellis:1975ap, Shifman:1979eb}, the couplings of the neutral fermions to the Higgs boson can be derived by the replacement $v \to v +h$ in the DM candidate mass $m_{\chi_1^0}(v)$:
\begin{equation}
\mathcal{L}_{h\Psi_1 \Psi_1} = \frac{1}{2} m_{\chi_1^0}(v+h) \bar\Psi_1 \Psi_1
	                         = \frac{1}{2} m_{\chi_1^0}(v) \bar\Psi_1 \Psi_1
										  +  \frac{1}{2} \frac{\partial m_{\chi_1^0}(v)}{\partial v} h \bar\Psi_1 \Psi_1 + \mathcal{O}(h^2),
\end{equation}
which means $C_{h,11}^\mathrm{S} = \partial m_{\chi_1^0}(v)/\partial v$~\cite{Cohen:2011ec, Cheung:2011aa}.

$m_{\chi_1^0}$ satisfies the characteristic equation
$\mathrm{det}(\mN - m_{\chi_1^0} \mathbbm{1}) =0$, which is just
\begin{equation}
	m_{\chi_1^0}^3 - m_S m_{\chi_1^0}^2 	-\frac{1}{2} (2 m_D^2 + y_1^2 v^2 + y_2^2 v^2) m_{\chi_1^0}
	+m_D (m_D m_S + y_1 y_2 v^2) =0.
\label{eq:eig}
\end{equation}
Differentiating its left-hand side with respect to $v$ and imposing $\partial m_{\chi_1^0}(v)/\partial v =0$, one obtain the condition that leads to $C_{h,11} =0$ is
\begin{equation}
	m_{\chi_1^0} = \frac{2 y_1 y_2 m_D}{y_1^2 + y_2^2}.
\end{equation}
Plugging this condition into Eq.~\eqref{eq:eig}, one obtains
\begin{equation}
	y_1 = \pm y_2\quad
\mathrm{or}\quad  y_1 = \frac{m_D \pm \sqrt{m_D^2 - m_S^2}}{m_S} y_2.
\label{eq:blind}
\end{equation}
Thus, the latter equation could also induce $C_{h,11}^\mathrm{S} =0$ when $m_D>m_S$.

\subsection{Higgs Precision Measurements at the CEPC}

\subsubsection{Corrections to the $Zh$ associated production}

The $Zh$ associated production $e^+e^-\to Zh$ is the primary Higgs production process in a Higgs factory with $\sqrt{s}=240-250~\si{GeV}$.
For the measurement of its cross section, a relative precision of $0.51\%$ is expected to be achieved at the CEPC with an integrated luminosity of $5~\iab$~\cite{CEPC-SPPCStudyGroup:2015csa}.
Below we discuss the impact of the SDFDM model on this cross section at one-loop level.

\begin{figure}[!tbp]
\centering
\includegraphics[width=.3\textwidth]{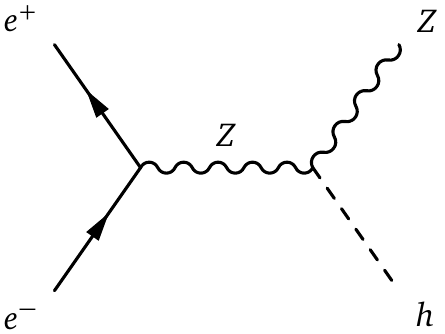}
\caption{Tree-level Feynman diagram for $e^+ e^- \to Z h$ in the SM.}
\label{fig:Born}
\end{figure}

\begin{figure}[!tbp]
\centering
\subfigure[\label{fig:v0}]
{\includegraphics[width=.3\textwidth]{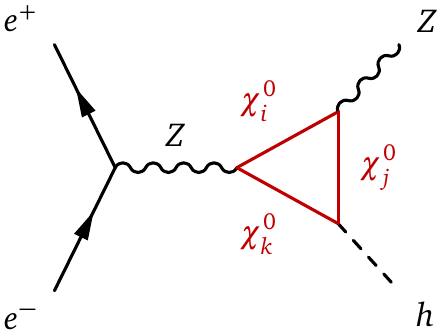}}
\hspace{.5em}
\subfigure[\label{fig:s0}]
{\includegraphics[width=.3\textwidth]{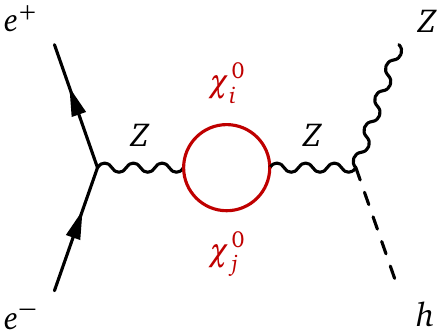}}
\hspace{.5em}
\subfigure[\label{fig:sc}]
{\includegraphics[width=.3\textwidth]{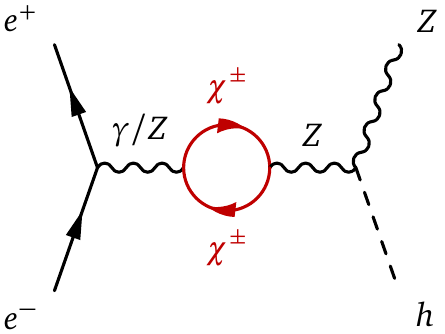}}
\caption{Feynman diagrams for vertex (a) and propagator (b, c) corrections to $e^+ e^- \to Z h$ due to the dark sector in the SDFDM model at one-loop level.}
\label{fig:v}
\end{figure}

\begin{figure}[!tbp]
\centering
\subfigure[\label{fig:Z_hh}]
{\includegraphics[height=.16\textwidth]{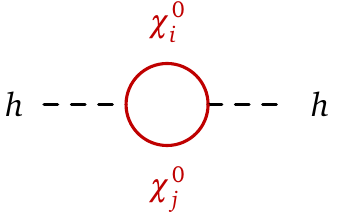}}
\hspace{1em}
\subfigure[\label{fig:Z_aZ}]
{\includegraphics[height=.16\textwidth]{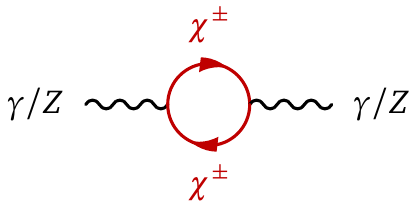}}\\
\subfigure[\label{fig:Z_ZZ}]
{\includegraphics[height=.16\textwidth]{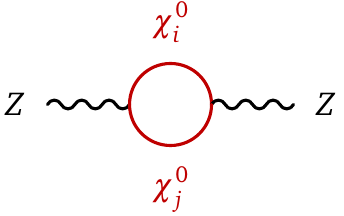}}
\hspace{1em}
\subfigure[\label{fig:Z_WW}]
{\includegraphics[height=.16\textwidth]{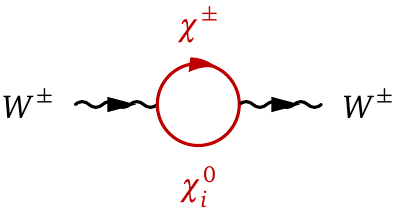}}
\caption{One-loop Feynman diagrams for self-energy corrections of the Higgs boson (a) and the electroweak gauge bosons (b, c, d) due to the dark sector in the SDFDM model.}
\label{fig:Z}
\end{figure}

Neglecting the extremely small $hee$ coupling, the only tree-level Feynman diagram for $e^+ e^- \to Z h$ in the SM is shown in Fig.~\ref{fig:Born}. It involves the $hZZ$ coupling, whose precise strength is a chief goal of a Higgs factory. BSM particles that couple to both the $Z$ and Higgs bosons, such as the Majorana fermions $\chi_i^0$, are presumed to modify this coupling via triangle loops, as demonstrated in Fig.~\ref{fig:v0}. Besides, Figs.~\ref{fig:s0} and \ref{fig:sc} show that dark sector fermions in the SDFDM model can also affect the propagator in the $e^+ e^- \to Z h$ diagram at one-loop level.
Moreover, as shown in Fig.~\ref{fig:Z}, the dark sector contributes to the self-energies of the Higgs boson and the electroweak gauge bosons, and hence influences the determination of the related renormalization constants. In practice, these contributions must be included to cancel the ultraviolet divergences from Fig.~\ref{fig:v}.

Formally, the $e^+e^-\to Zh$ cross section can be split into two parts:
\begin{equation}
	\sigma = \sigma_0 + \sigma_\mathrm{BSM},
\end{equation}
where $\sigma_0$ is the SM prediction, while $\sigma_\mathrm{BSM}$ is the contribution due to BSM physics, which, in our case, is the dark sector multiplets. The next-to-leading corrections to $e^+e^- \to Zh$ in the SM have been calculated  two decades ago~\cite{Fleischer:1982af,Kniehl:1991hk,Denner:1992bc,DENNER1992263}, while the mixed electroweak-QCD ($\mathcal{O}(\alpha\alpha_\mathrm{s})$) corrections have been studied in 2016~\cite{Gong:2016jys,Sun:2016bel}.
Here we calculate $\sigma_0$ with one-loop corrections except for the virtual photon correction. Thus, we would not need to involve the real photon radiation process $e^+e^-\to Zh\gamma$ for dealing with soft and collinear divergences. This treatment should be sufficient for our purpose, as we are only interested in the relative deviation of the $e^+e^-\to Zh$ cross section due to the dark sector.

\begin{figure}[!tbp]
\centering
\includegraphics[width=.5\textwidth]{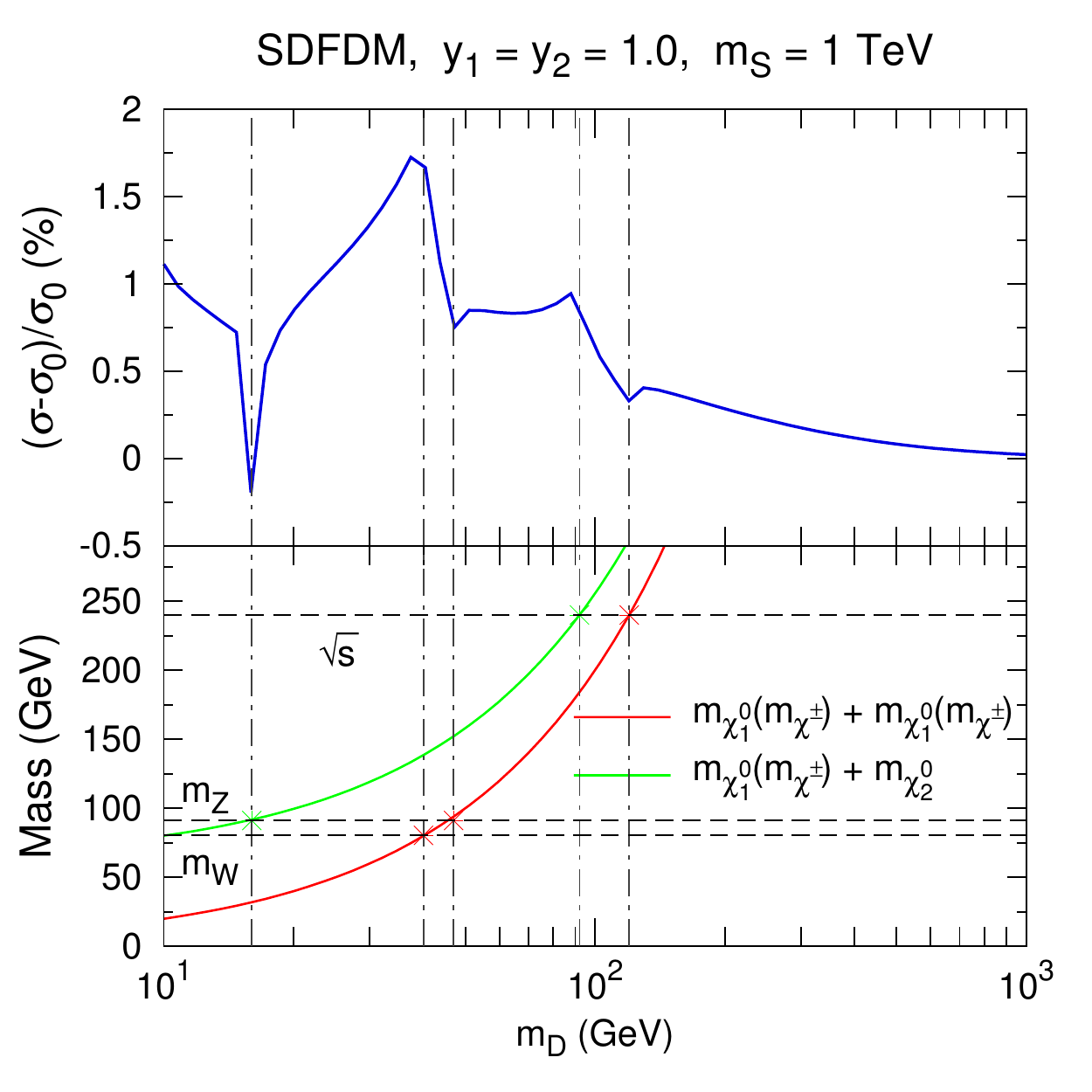}
\caption{Relative deviation of the $e^+e^-\to Zh$ cross section at $\sqrt{s} = 240~\GeV$ in the SDFDM model.
The lower frame shows the sums of dark sector fermion masses in order to demonstrate threshold effects with $m_Z = \mchia + \mchib$, $m_W = \mchia + \mchipm$, $m_Z = 2\mchipm$, and $\sqrt{s} = \mchia + \mchib$.}
\label{fig:12eehz}
\end{figure}

We utilize the packages \texttt{FeynArts 3.9}~\cite{Hahn:2000kx}, ~\texttt{FormCalc 9.4}~\cite{Hahn:1998yk}, and ~\texttt{LoopTools 2.13}~\cite{vanOldenborgh:1990yc} to calculate one-loop corrections from the SM and from the SDFDM model at $\sqrt{s}=240~\si{GeV}$.
The on-shell renormalization scheme is adopted to fix the renormalization constants.
Fig.~\ref{fig:12eehz} shows the relative deviation of the $e^+e^-\to Zh$ cross section $(\sigma-\sigma_0)/\sigma_0$ as a function of $m_D$. Other parameters are chosen to be $y_1=y_2=1$ and $m_S=1~\si{TeV}$, leading to $\mchipm=\mchia$. The deviation could be either positive or negative, depending on the parameters. As $m_D$ increases to the TeV scale, the deviation becomes very small, because the dark sector basically decouples.

When the dark sector fermions in the loops are able to close to their mass shells, their contributions could vary dramatically.
In the lower frame of Fig.~\ref{fig:12eehz} shows the sums of fermion masses in order to demonstrate the mass threshold effects with  $m_Z = \mchia + \mchib$, $m_W = \mchia + \mchipm$, $m_Z = 2\mchipm$, and $\sqrt{s} = \mchia + \mchib$.
For instance, $m_Z > \mchia + \mchib$ would allow a new decay process, $Z\to\chia\chib$; this means that the $Z$ boson self-energy develops a new imaginary part, which is absent for $m_Z < \mchia + \mchib$. As a result, $(\sigma-\sigma_0)/\sigma_0$ reaches a dip at $m_Z = \mchia + \mchib$.
Similarly, we have threshold effects with $m_W = \mchia + \mchipm$ and $m_Z = 2\mchipm$.
In addition, the threshold effect with $\sqrt{s} = \mchia + \mchib$ is caused by the triangle loop in Fig.~\ref{fig:v0}, because $\sqrt{s} > \mchia + \mchib$ also leads to a imaginary part in the amplitude of the triangle loop.

\begin{figure}[!tbp]
\centering
\subfigure[~$y_1 = 0.5$, $y_2 = 1.5$.\label{fig:12zh:a}]
{\includegraphics[width=.45\textwidth,trim={0 30 0 10},clip]{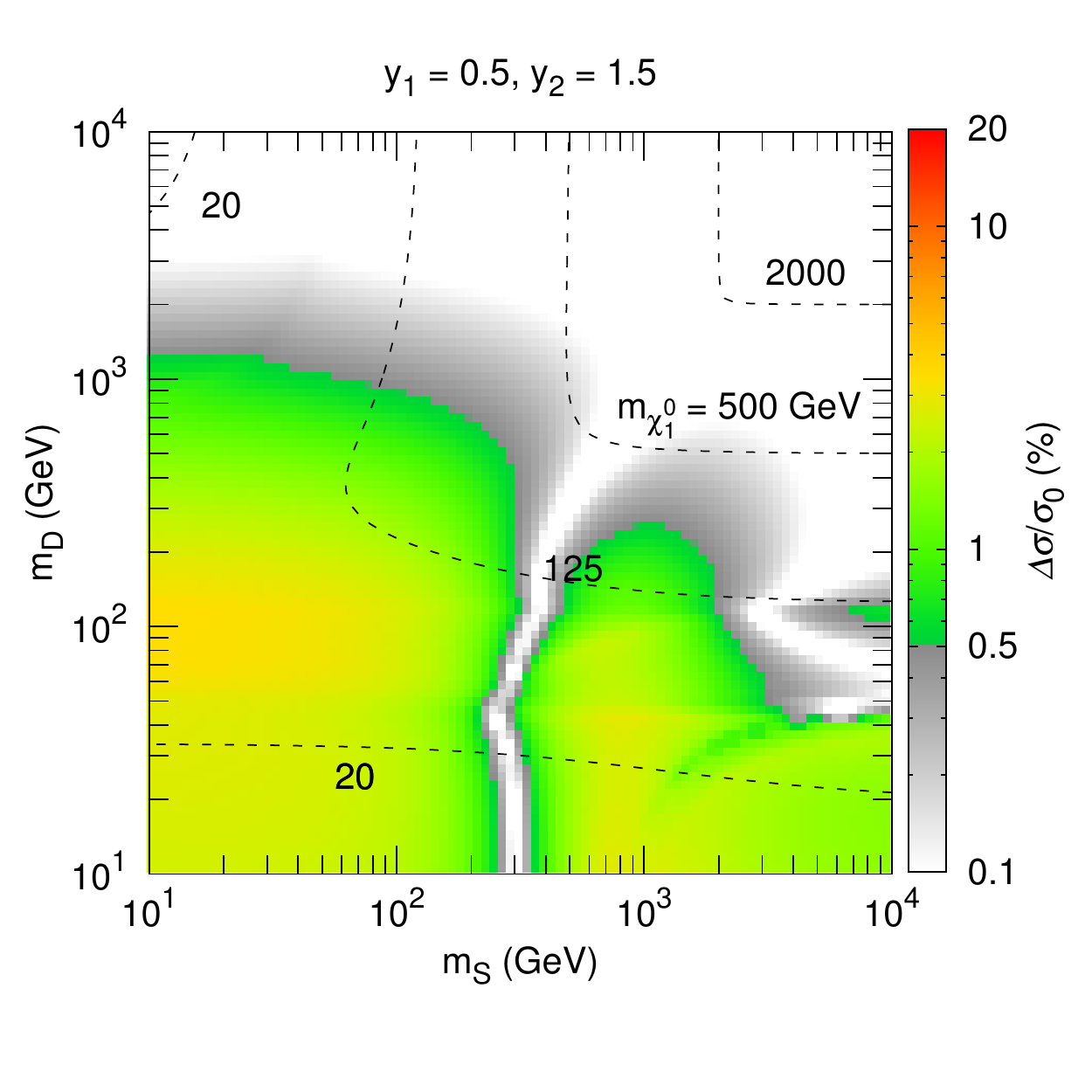}}
\subfigure[~$y_1 = y_2 = 1.0$.\label{fig:12zh:b}]
{\includegraphics[width=.45\textwidth,trim={0 30 0 10},clip]{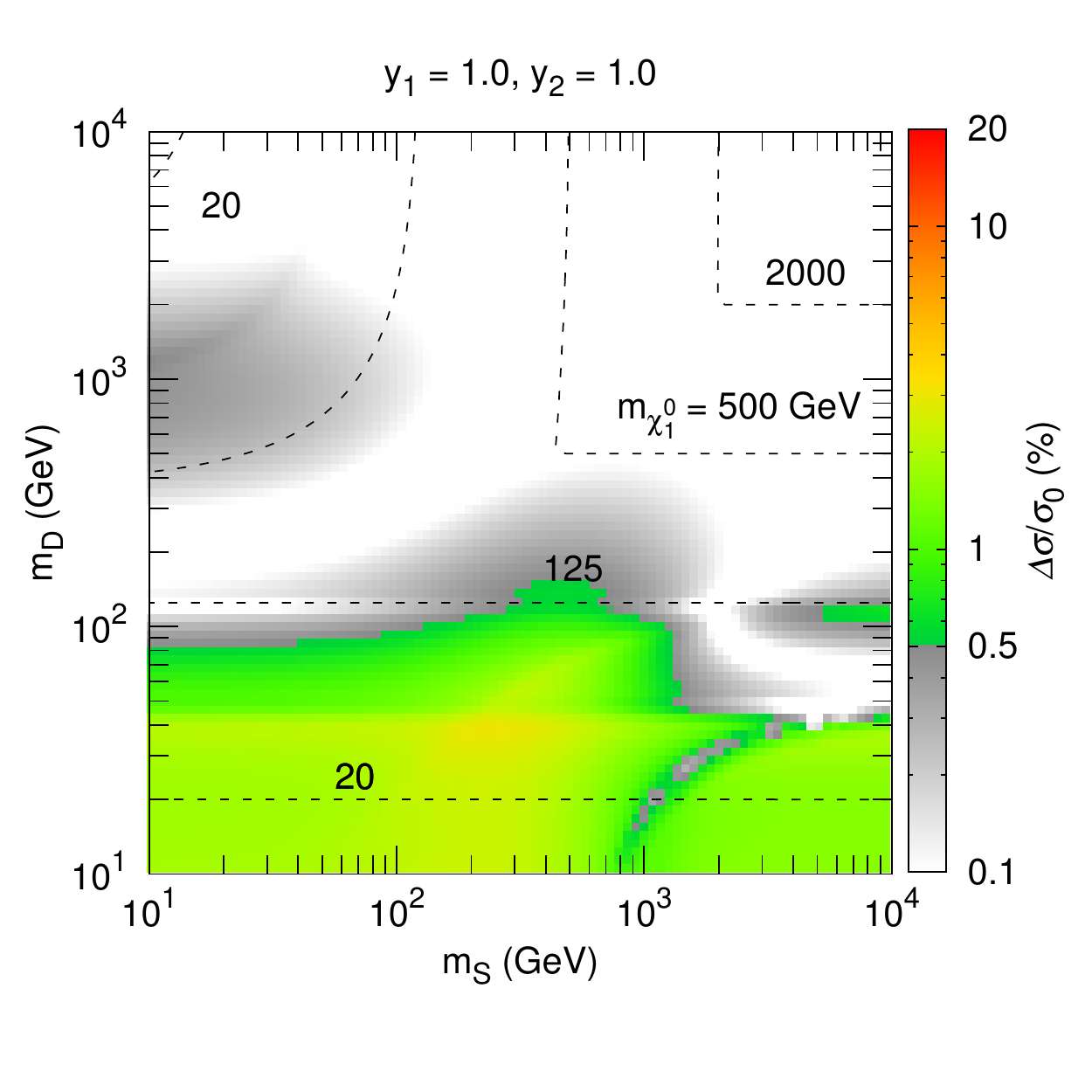}}
\subfigure[~$m_S = 100~\GeV$, $m_D = 400~\GeV$.\label{fig:12zh:c}]
{\includegraphics[width=.45\textwidth,trim={0 30 0 10},clip]{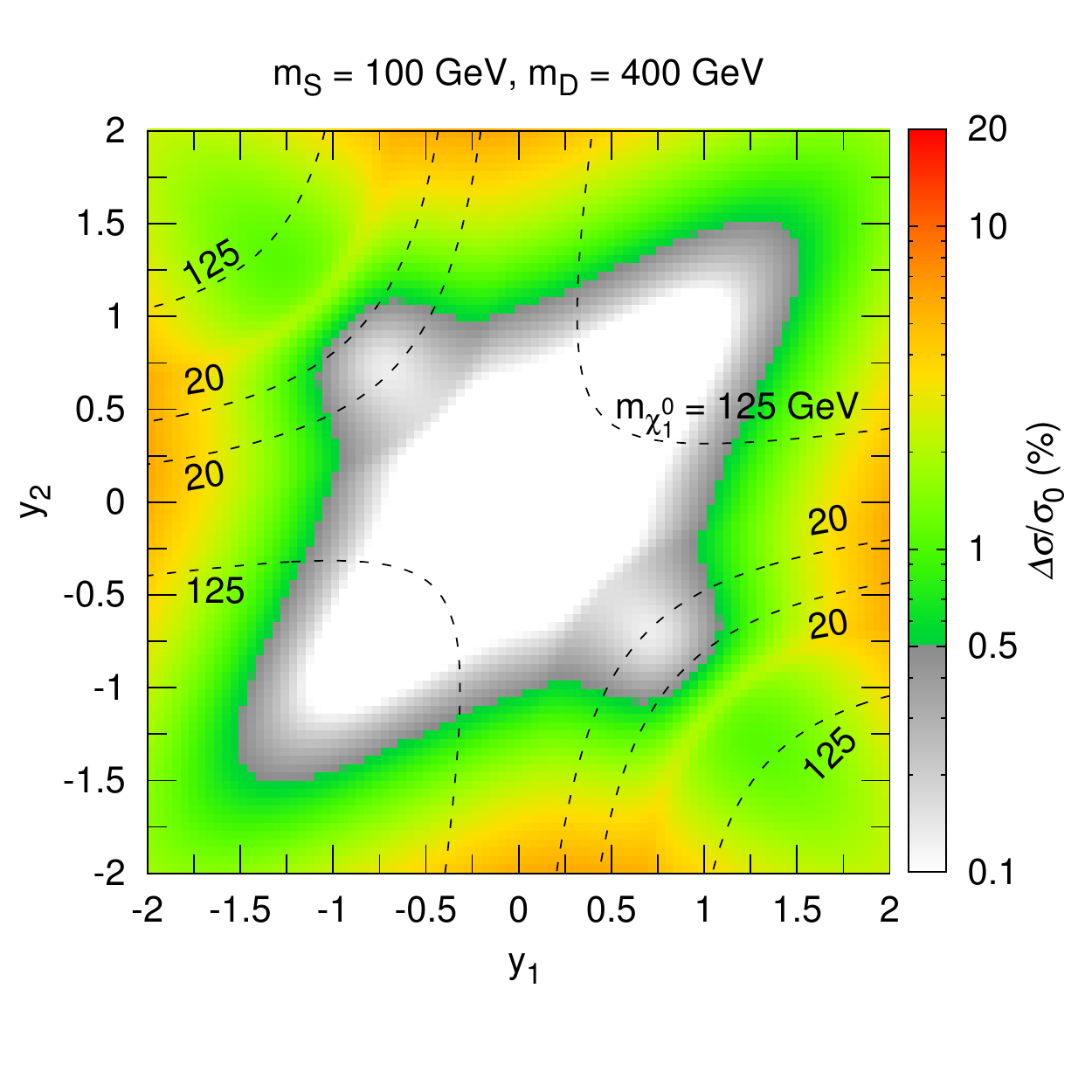}}
\subfigure[~$m_S = 400~\GeV$, $m_D = 150~\GeV$.\label{fig:12zh:d}]
{\includegraphics[width=.45\textwidth,trim={0 30 0 10},clip]{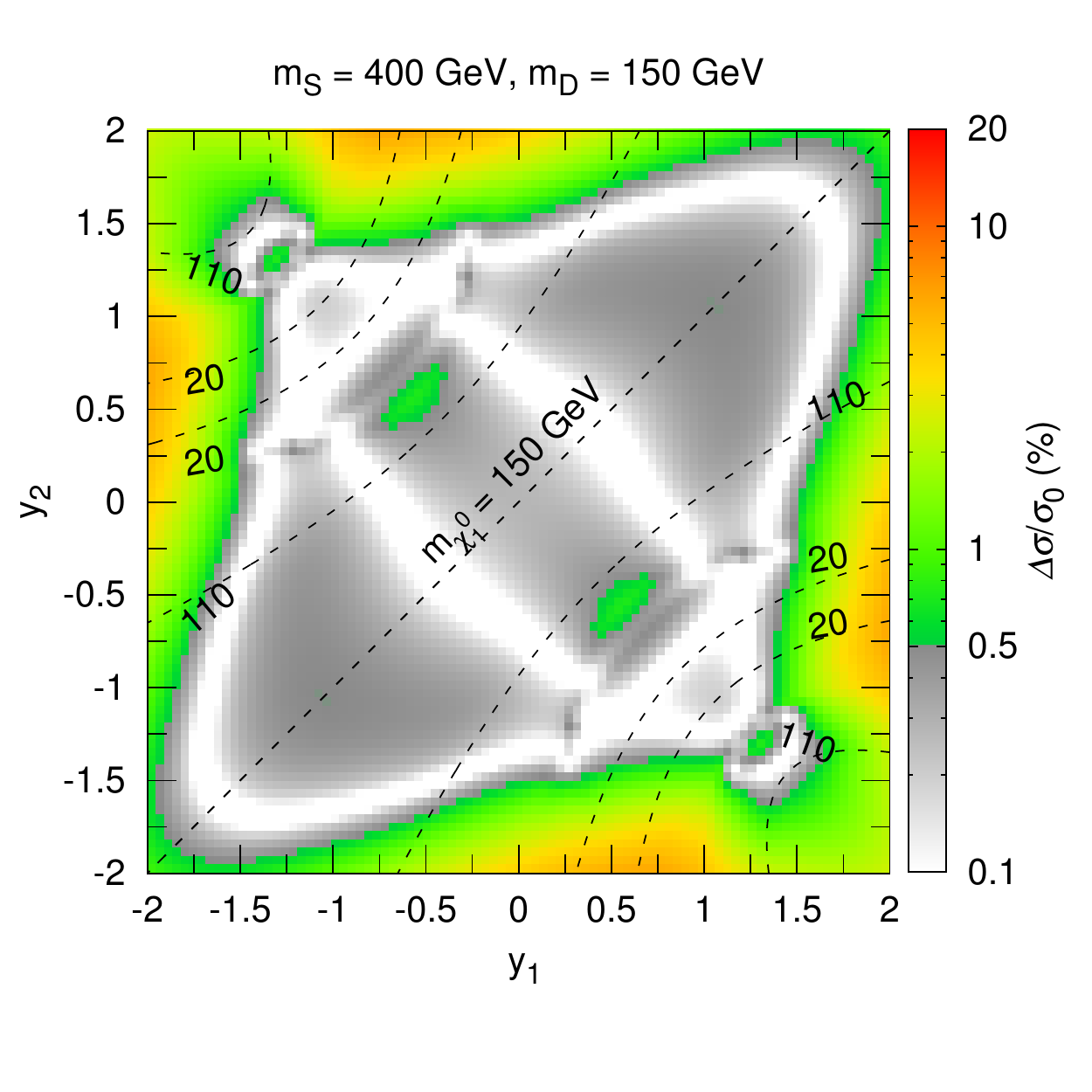}}
\caption{Heat maps for the absolute relative deviation of the $e^+e^-\to Zh$ cross section $\Delta \sigma/\sigma_0 \equiv |\sigma - \sigma_0|/\sigma_0$ in the SDFDM model. Results are shown in the $m_S-m_D$ (a,b) and $y_1-y_2$ (c,d) planes with two parameters fixed as indicated.
Colored and gray regions correspond to $\Delta \sigma/\sigma_0 > 0.5\%$ and $< 0.5\%$, respectively.
Dashes lines denote contours of the DM candidate mass $\mchia$.}
\label{fig:12zh}
\end{figure}

In Fig.~\ref{fig:12zh}, we show heat maps for the absolute relative deviation $\Delta \sigma/\sigma_0 \equiv |\sigma - \sigma_0|/\sigma_0$ in the SDFDM model with two parameters fixed.
The regions with colors have sufficient deviations that could be explored by the CEPC measurement of the $e^+e^-\to Zh$ cross section, while the gray regions are beyond its capability. The complicated behaviors of these heat maps can be attributed to
mass threshold effects, as shown in Fig.~\ref{fig:12eehz}.

For $y_1=0.5$ and $y_2=1.5$ [Fig.~\ref{fig:12zh:a}], the CEPC measurement could probe up to $\mchia \sim 200~\si{GeV}$. For $y_1=y_2=1$  [Fig.~\ref{fig:12zh:a}], where the custodial symmetry is respected, regions with $\mchia \gtrsim m_h$ could hardly have apparent deviations.
Furthermore, Figs.~\ref{fig:12zh:c} and \ref{fig:12zh:d} show that larger Yukawa couplings $y_1$ and $y_2$ basically induce larger $\Delta \sigma/\sigma_0$ for fixed $m_S$ and $m_D$.

\subsubsection{Higgs boson invisible decay}

If the dark sector fermions are sufficient light, the Higgs boson and the $Z$ boson would be able to decay into them.
When such decay processes are kinematically allowed, their widths are given by ($i\neq j$ in the expressions below)
\begin{eqnarray}
\Gamma (h\to\chi_i^0\chi_j^0) &=& \frac{{F(m_h^2,m_{\chi _i^0}^2,m_{\chi _j^0}^2)}}{{32\pi m_h^3}}
\big\{ |C_{h,ij}^{\mathrm{S}} + C_{h,ji}^{\mathrm{S}}{|^2}[m_h^2 - {({m_{\chi _i^0}} + {m_{\chi _j^0}})^2}] \nonumber\\
&& + |C_{h,ij}^{\mathrm{P}} + C_{h,ji}^{\mathrm{P}}{|^2}[m_h^2 - {({m_{\chi _i^0}} - {m_{\chi _j^0}})^2}]\big\},\label{eq:h2chi0ichi0j}\\
\Gamma (h\to\chi_i^0\chi_i^0) &=& \frac{{|C_{h,ii}^{\mathrm{S}}{|^2}}}{{16\pi m_h^2}}{(m_h^2 - 4m_{\chi _i^0}^2)^{3/2}},\label{eq:h2chi0ichi0i}\\
\Gamma (Z\to\chi_i^0\chi_j^0) &=& \frac{{F(m_Z^2,m_{\chi _i^0}^2,m_{\chi _j^0}^2)}}{{24\pi m_Z^5}}\big\{ 6(|C_{Z,ij}^{\mathrm{V}}{|^2} - |C_{Z,ij}^{\mathrm{A}}{|^2})m_Z^2{m_{\chi _i^0}}{m_{\chi _j^0}} \nonumber\\
&& + (|C_{Z,ij}^{\mathrm{A}}{|^2} + |C_{Z,ij}^{\mathrm{V}}{|^2})[m_Z^2(2m_Z^2 - m_{\chi _i^0}^2 - m_{\chi _j^0}^2) - {(m_{\chi _i^0}^2 - m_{\chi _j^0}^2)^2}]\big\},\label{eq:Z2chi0ichi0j}\\
\Gamma (Z\to\chi_i^0\chi_i^0) &=& \frac{{|C_{Z,ii}^{\mathrm{A}}{|^2}}}{{24\pi m_Z^2}}{(m_Z^2 - 4m_{\chi _i^0}^2)^{3/2}},\label{eq:Z2chi0ichi0i}\\
\Gamma (Z\to\chi^+\chi^-) &=& \frac{{{g^2}{{(c_{\mathrm{W}}^2 - s_{\mathrm{W}}^2)}^2}}}{{48\pi m_Z^2c_{\mathrm{W}}^2}}\sqrt {m_Z^2 - 4m_{{\chi ^ + }}^2} (m_Z^2 + 2m_{{\chi ^ + }}^2),
\end{eqnarray}
where $F(x,y,z) \equiv \sqrt {{x^2} + {y^2} + {z^2} - 2xy - 2xz - 2yz}$.

Since $\chia$ cannot be directly probed by detectors in collider experiments, the decay processes $h\to\chia\chia$ and $Z\to\chia\chia$ are invisible. On the other hand, if $h$ and $Z$ decay into other dark sector fermions, the $Z_2$ symmetry will force them subsequently decay into $\chia$ associated with SM particles in final states. Such $h$ and $Z$ decays may also be invisible due to $\chi_{2,3}^0\to \chi_1^0 Z^*(\to \nu\bar\nu)$. Moreover, when these decay processes are allowed, the SM products would probably be very soft, as the related mass spectrum in the dark sector should be compressed. As a result, they could be effectively invisible.
Therefore, the invisible decays of $h$ and $Z$ provide another promising approach to reveal the dark sector.

With an integrated luminosity of $5~\iab$, CEPC is expected to constrain the branching ratio of the invisible decay down to $0.28\%$ at 95\% CL~\cite{CEPC-SPPCStudyGroup:2015csa}.
As the Higgs boson width in the SM is 4.08~MeV~ for $m_h=125.1~\si{GeV}$~\cite{Heinemeyer:2013tqa}, this means that the expected constraint on the Higgs invisible decay width is $\Gamma_{h,\mathrm{inv}}<11.4~\keV$.
On the other hand, LEP experiments have put an upper bound on the $Z$ invisible width, which is $\Gamma_{Z,\mathrm{inv}}^\mathrm{BSM} < 2~\MeV$ at $95\%$ CL~\cite{ALEPH:2005ab}.

\begin{figure}[!tbp]
\centering
\subfigure[~$y_1 = 0.5$, $y_2 = 1.5$.]
{\includegraphics[width=.45\textwidth,trim={0 15 0 10},clip]{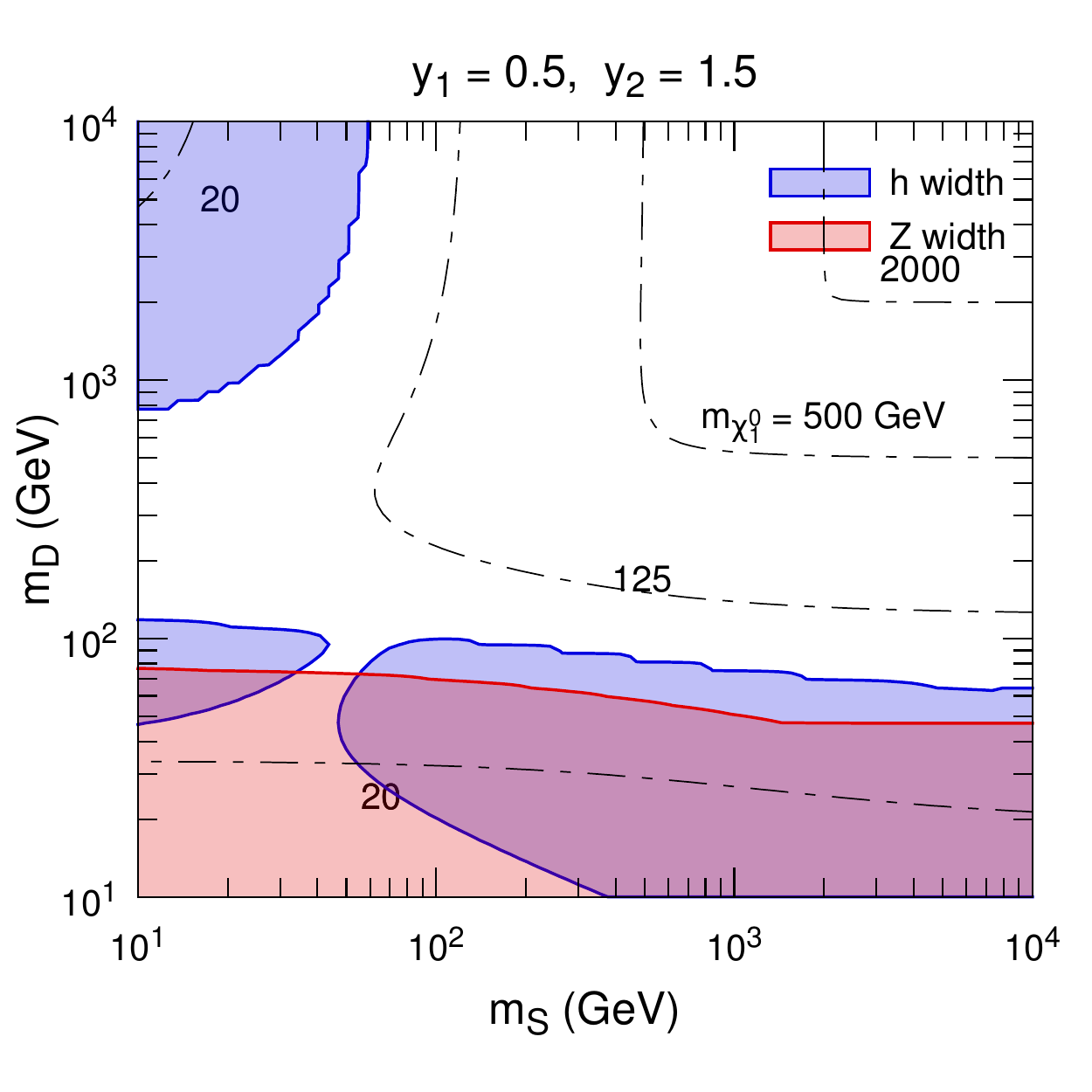}}
\subfigure[~$y_1 = y_2 = 1.0$.]
{\includegraphics[width=.45\textwidth,trim={0 15 0 10},clip]{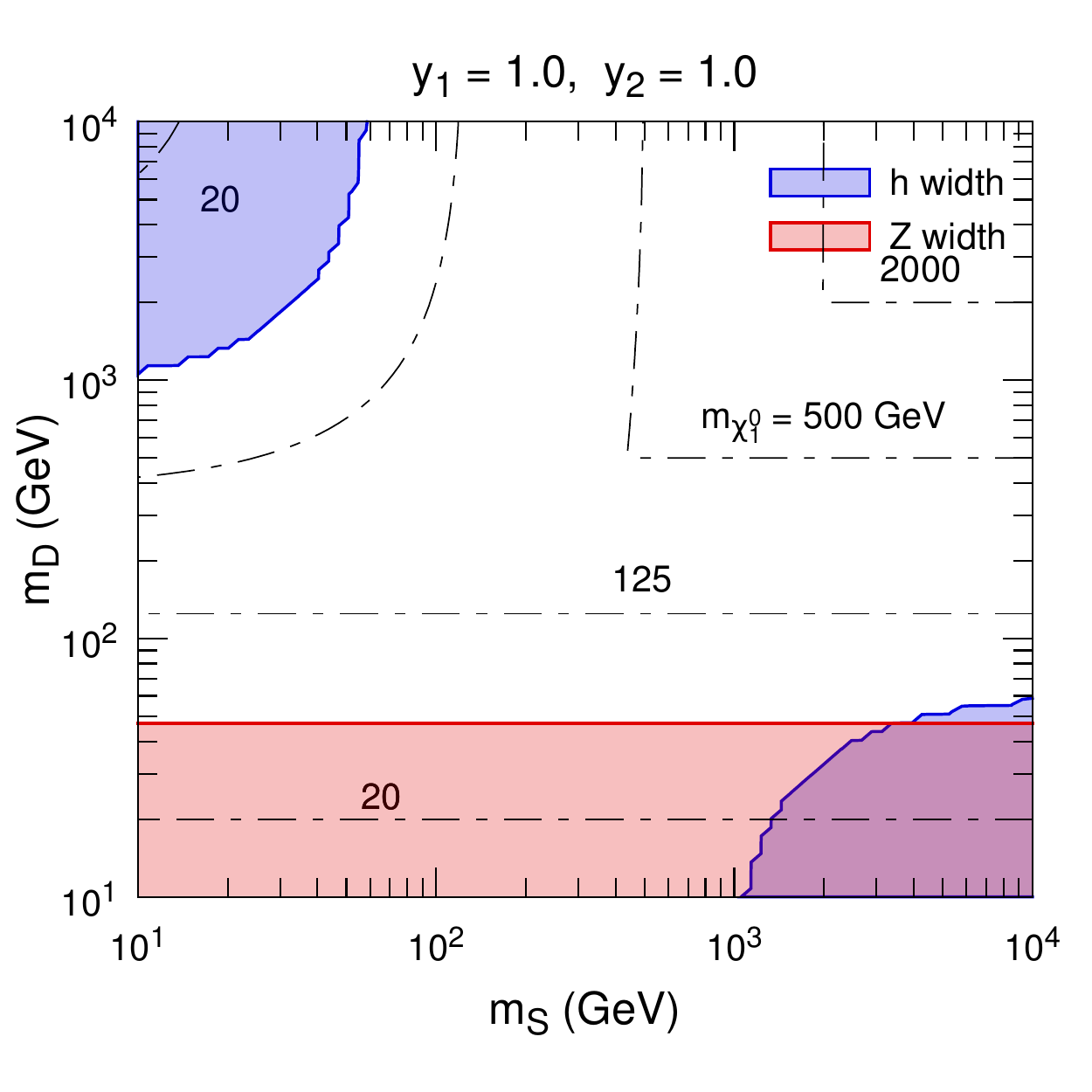}}
\subfigure[~$m_S = 100~\GeV$, $m_D = 400~\GeV$.]
{\includegraphics[width=.45\textwidth,trim={0 15 0 10},clip]{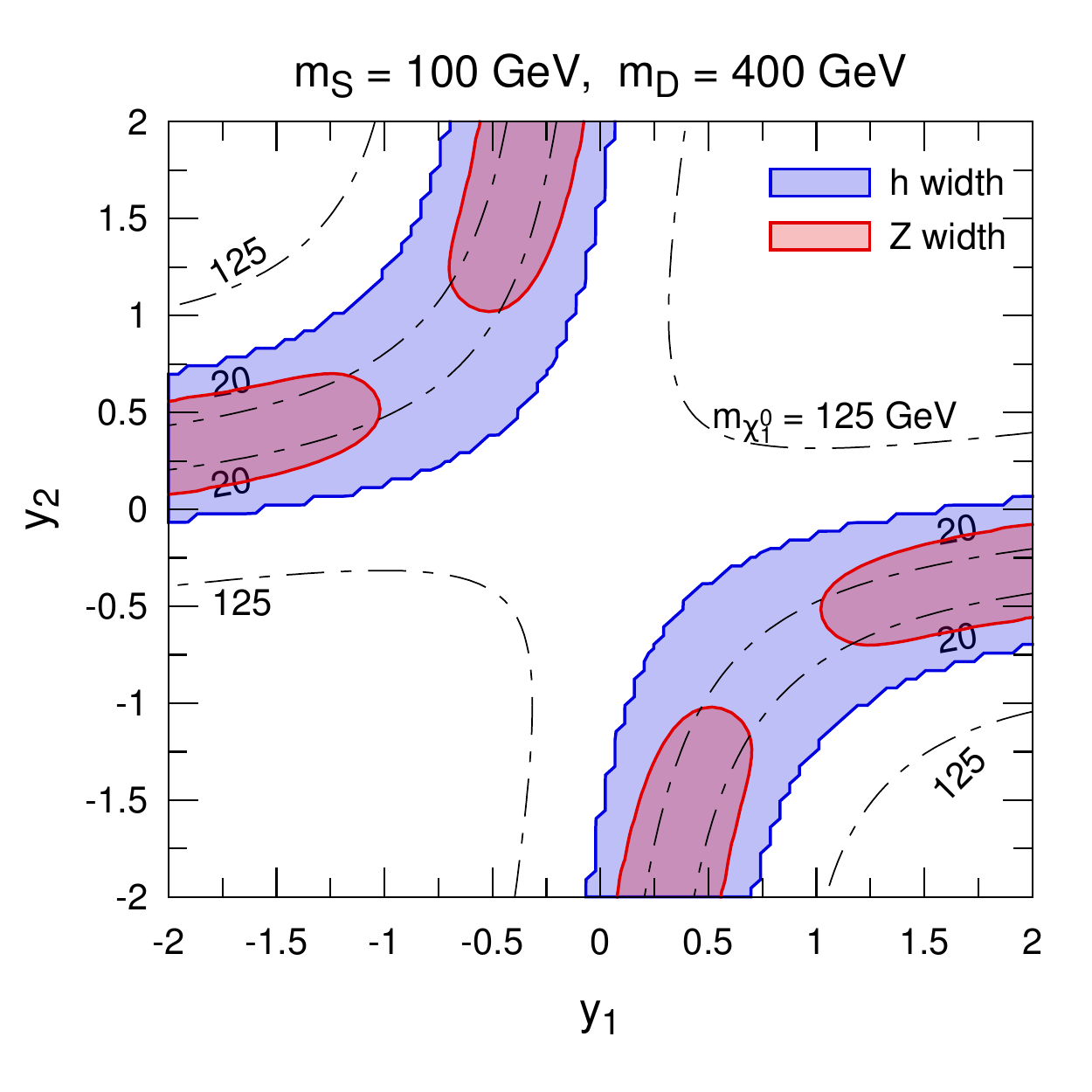}}
\subfigure[~$m_S = 400~\GeV$, $m_D = 150~\GeV$.]
{\includegraphics[width=.45\textwidth,trim={0 15 0 10},clip]{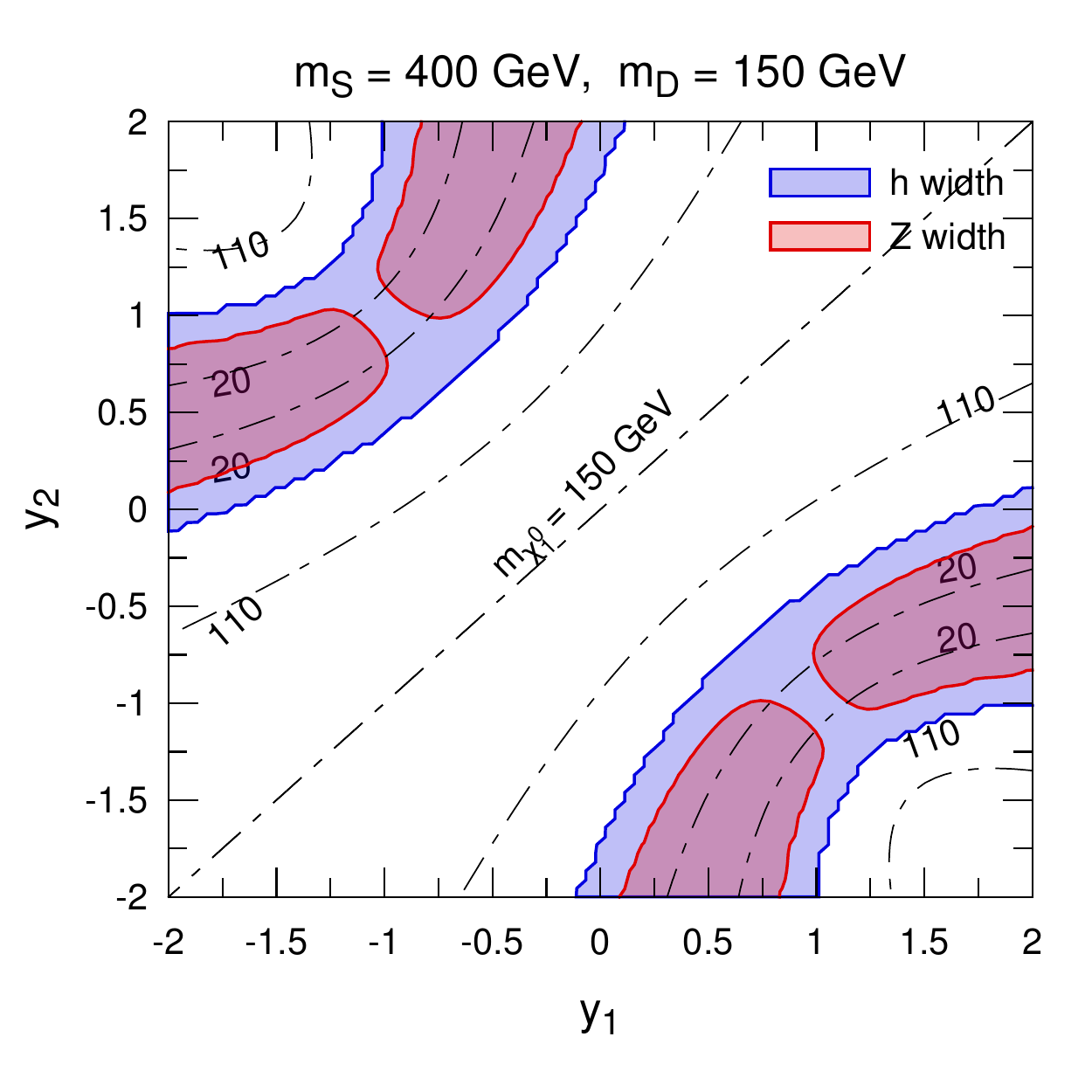}}
\caption{95\% CL expected constraints (blue regions) from the CEPC measurement of the Higgs boson invisible decay width in the $m_S-m_D$ plane (a,b) and the $y_1-y_2$ plane (c,d) for the SDFDM model.
Red regions have been excluded at 95\% CL by the measurement of the $Z$ boson invisible decay width in LEP experiments~\cite{ALEPH:2005ab}.
Dot-dashed lines indicate $m_{\chi_1^0}$ contours.
}
\label{fig:12width}
\end{figure}

In Fig.~\ref{fig:12width} we present the expected CEPC constraint and the LEP constraint from the invisible decays of the Higgs boson and the $Z$ boson, respectively.
We have included all allowed decay channels into the dark sector as invisible decays for the reasons we mentioned above. Although this treatment overestimates the invisible decay widths, it actually closes to the most conservative estimation that only takes into account $h\to\chia\chia$ and $Z\to\chia\chia$, because in most of the parameter regions we are interested in only one or a few of these decay channels would open.
From Fig.~\ref{fig:12width}, we can see that the expected CEPC constraints from the Higgs invisible decay are basically stronger than the LEP constraint from the $Z$ invisible decay.
Exceptions happen mostly when $m_D < m_{h}/2$. In such a region, the $Z\to\chi^+\chi^-$ decay is allowed, while the $C_{h,11}^\mathrm{S}$ coupling for $m_D<m_S$ could be small, or even vanishes if $y_1=y_2$.

\subsection{Current experimental constraints}

In this subsection, we investigate current experimental constraints on the SDFDM model.
Relevant bounds come from the observation of DM relic abundance, DM direct detection experiment, LHC monojet searches, and LEP searches for charged particles.
Below we discuss them one by one.

\subsubsection{Relic abundance}

The observed cold DM relic density reported by the Planck collaboration is $\Omega_\mathrm{DM} h^2 = 0.1186\pm 0.0020$~\cite{Planck:2015xua}.
Assuming DM particles were thermally produced in the early Universe, the relic density is determined by their thermally averaged annihilation cross section into SM particles when they decoupled.
If the annihilation cross section is too small, DM would be overproduced, contradicting the observation.

The freeze-out temperature is controlled by the DM particle mass, which is $m_{\chia}$ in the SDFDM model.
However, other dark sector fermions may have masses similar to $m_{\chia}$.
For instance, $m_S > m_D$ could lead to a doublet-dominated $\chia$, whose mass can be very close to $m_{\chi^\pm}$ and $m_{\chib}$.
As a result, coannihilation processes among the dark sector fermions could be important and significantly influence the DM relic abundance.
For this reason, we take into account the coannihilation effect when the mass differences are within $0.1 m_{\chi_1^0}$.
We adopt \texttt{MadDM}~\cite{Backovic:2013dpa}, which is based on \texttt{MadGraph 5}~\cite{Alwall:2014hca}, to calculate the relic density involving all annihilation and coannihilation channels.
The model is implemented with \texttt{FeynRules 2}~\cite{Alloul:2013bka}.

\begin{figure}[!tbp]
\centering
\subfigure[~$y_1 = 0.5$, $y_2 = 1.5$.\label{fig:12cons:a}]
{\includegraphics[width=.45\textwidth,trim={0 15 0 10},clip]{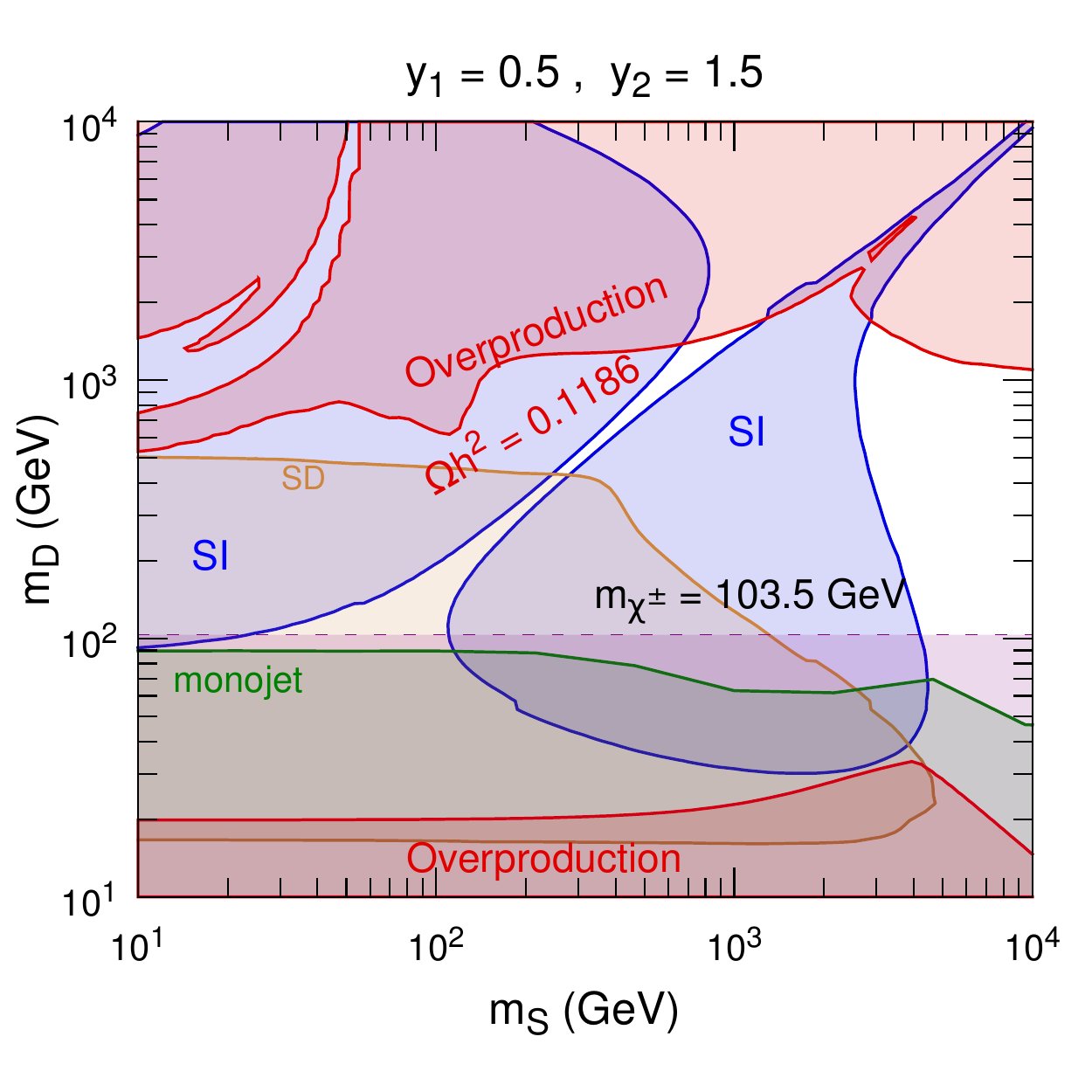}}
\subfigure[~$y_1 = y_2 = 1.0$.\label{fig:12cons:b}]
{\includegraphics[width=.45\textwidth,trim={0 15 0 10},clip]{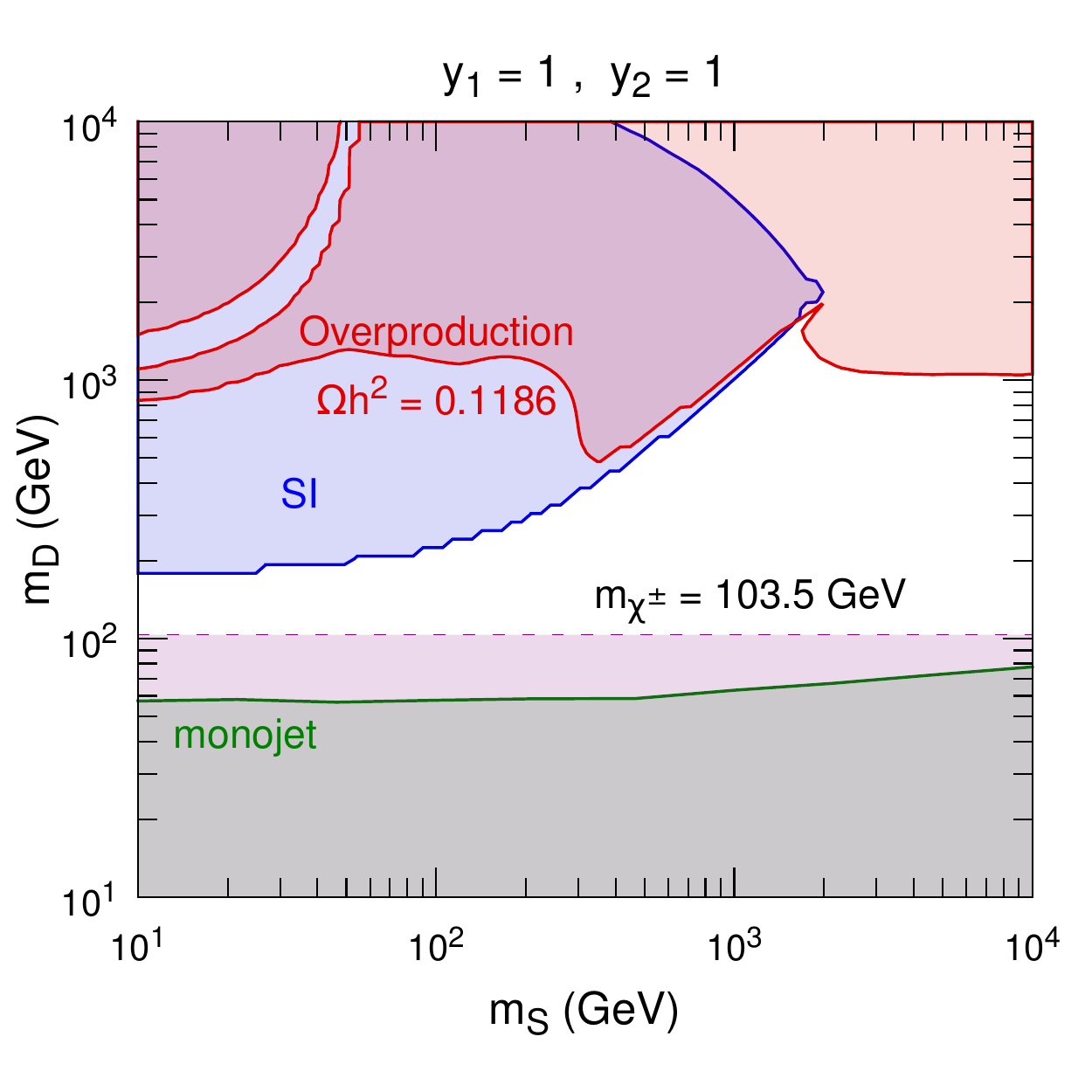}}
\subfigure[~$m_S = 100~\GeV$, $m_D = 400~\GeV$.\label{fig:12cons:c}]
{\includegraphics[width=.45\textwidth,trim={0 15 0 10},clip]{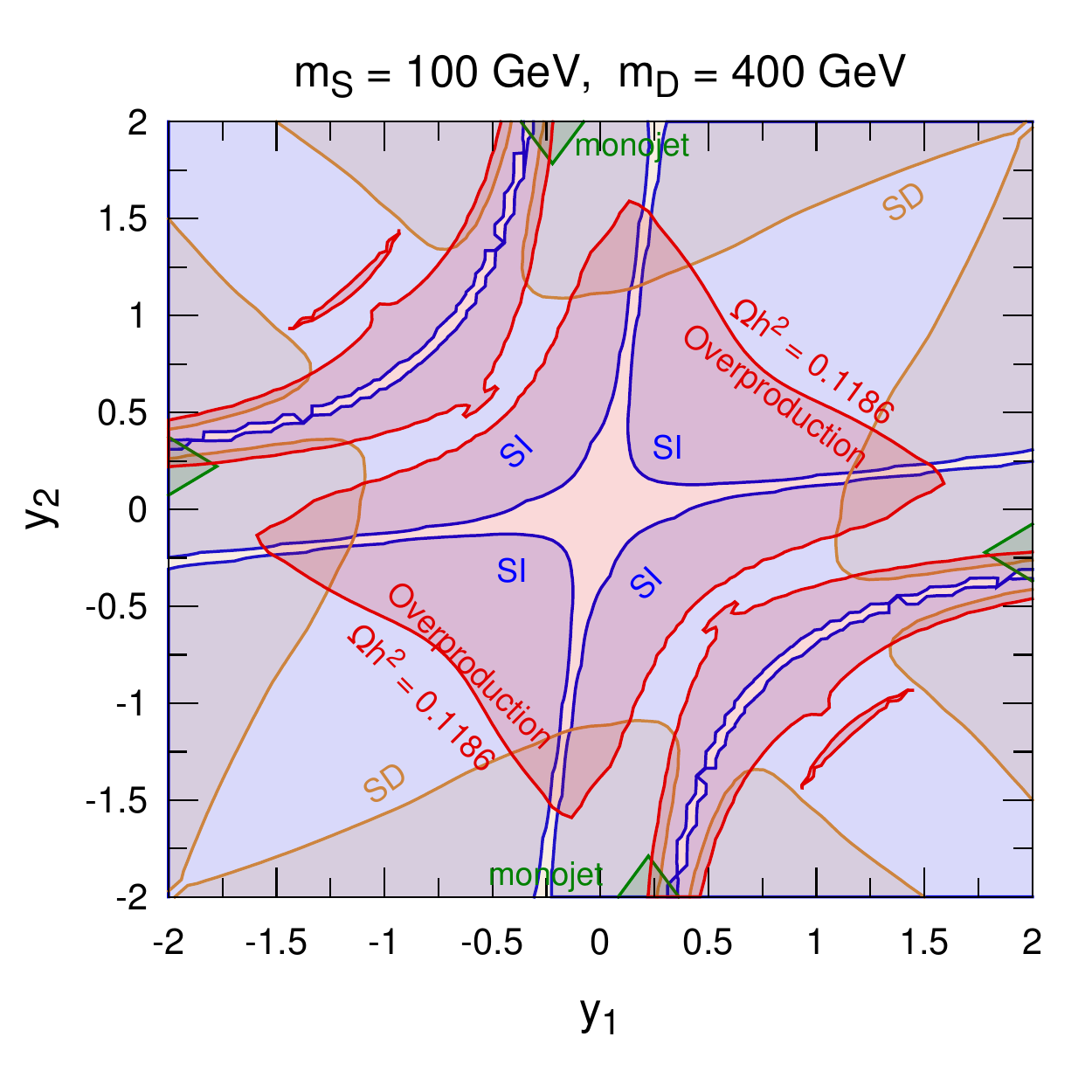}}
\subfigure[~$m_S = 400~\GeV$, $m_D = 150~\GeV$.\label{fig:12cons:d}]
{\includegraphics[width=.45\textwidth,trim={0 15 0 10},clip]{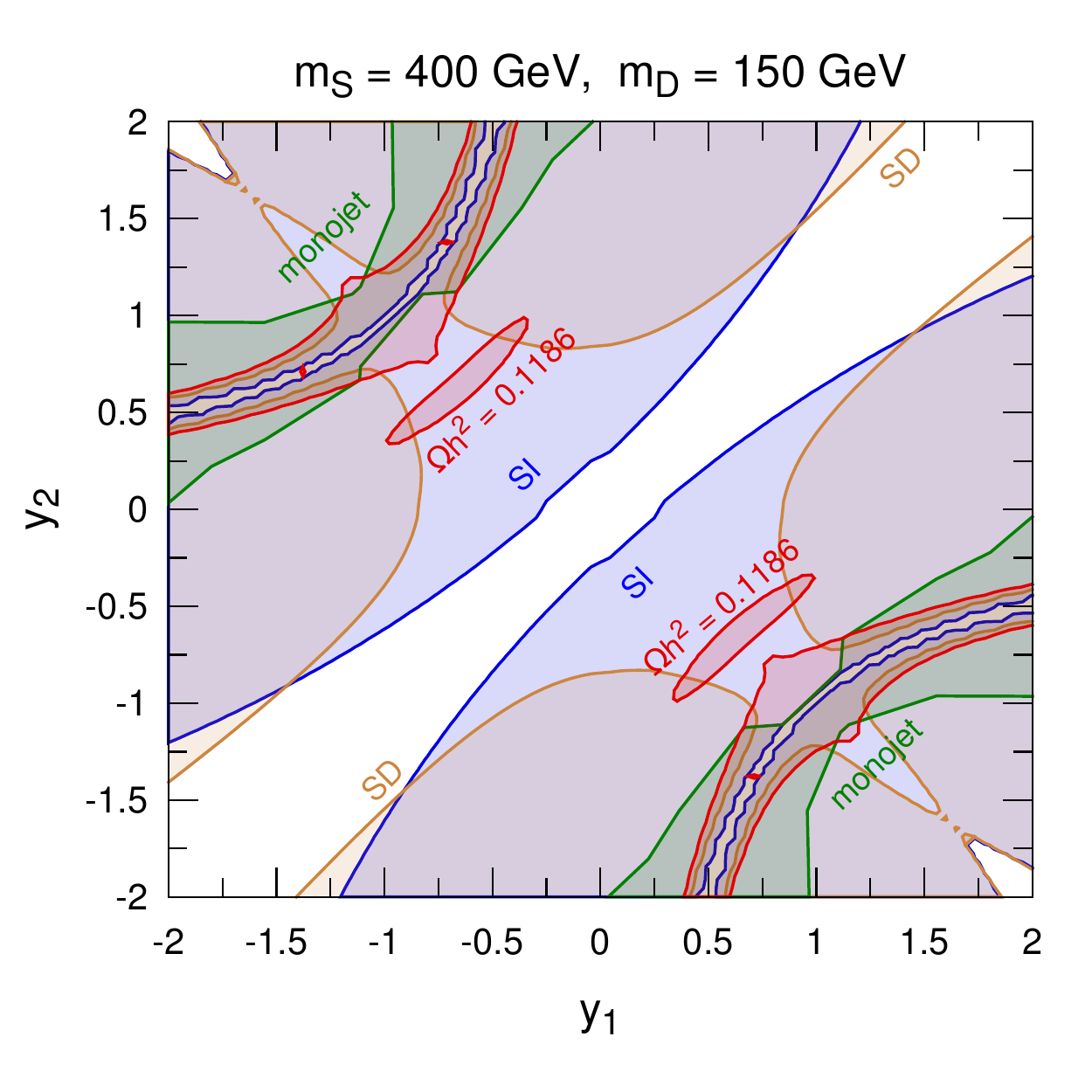}}
\caption{Experimental constraints in the $m_S-m_D$ plane (a,b) and the $y_1-y_2$ plane (c,d) for the SDFDM model.
The red regions indicate DM overproduction in the early Universe.
The blue and orange regions are excluded by the PandaX direct detection experiment for SI interactions~\cite{Tan:2016zwf} and for SD interactions~\cite{Fu:2016ega}, respectively.
The green regions are ruled out by the ATLAS monojet search~\cite{Aad:2015zva}.
The pink regions are excluded by the search for charged particles at the LEP~\cite{Abdallah:2003xe}.}
\label{fig:12cons}
\end{figure}

The parameter regions where DM is overproduced are indicated by red color in Fig.~\ref{fig:12cons}.
For a DM candidate purely from the doublets, the observed relic abundance corresponds to a DM particle mass of $\sim 1.2~\si{TeV}$~\cite{Cirelli:2005uq}. The mixing with the singlet complicates the situation. Nonetheless, Figs.~\ref{fig:12cons:a} and \ref{fig:12cons:b} still show that the observation favors $m_D\sim \si{TeV}$.
Annihilation through a $Z$ or $h$ resonance would significantly increase the cross section and hence reduce the relic density. This effect results in the bands of underproduction among the overproduction regions in Figs.~\ref{fig:12cons:a} and \ref{fig:12cons:b}.

Fig.~\ref{fig:12cons:a} also has a overproduction region with $m_D \lesssim 30~\si{GeV}$, due to lacking of effective annihilation mechanisms.
In this region, $m_{\chia} \lesssim 30~\si{GeV}$ forbids the annihilation into weak gauge bosons, while the annihilation into SM fermions is helicity-suppressed and the coannihilation effect with $\chipm$ is insufficient.
A similar region dose not show up in Fig.~\ref{fig:12cons:b}, because in this case $m_{\chia}=m_{\chipm}=m_D$ leads to a significant coannihilation effect.
Figs.~\ref{fig:12cons:c} and \ref{fig:12cons:d} demonstrate the complicate overproduction regions depending on the Yukawa couplings for specified mass parameters of the dark sector.

\subsubsection{DM direct detection}

The $Z\chia\chia$ and $h\chia\chia$ couplings could induce spin-dependent (SD) and spin-independent (SI) DM-nucleon scattering, respectively. Therefore, the model is testable in direct detection experiments.
\texttt{MadDM}~\cite{Backovic:2015cra} is used to calculate the DM-nucleon scattering cross sections.
We also present the results in Fig.~\ref{fig:12cons}, with blue and orange regions excluded at 90\% CL by the PandaX experiment for SI interactions~\cite{Tan:2016zwf} and for SD interactions~\cite{Fu:2016ega}, respectively.

As in this model SI and SD interactions have different origins, their effects are comparable and complementary in direct detection experiments, as shown in Fig.~\ref{fig:12cons:a}, \ref{fig:12cons:c}, and \ref{fig:12cons:d}.
When $y_1 = y_2$, the $Z\chia\chia$ coupling vanishes, and thus there is no SD exclusion region in Fig.~\ref{fig:12cons:b}.
Moreover, as $y_1 = y_2$ and $m_S > m_D$ lead to a vanishing $h\chia\chia$ coupling, no SI constraint is available in the related regions of Figs.~\ref{fig:12cons:b} and Fig.~\ref{fig:12cons:d}.

In Fig.~\ref{fig:12cons:c}, the model is severely constrained by DM direct detection.
Exceptions occur when the $h\chia\chia$ coupling happens to vanish.
For $m_S = 100~\GeV$ and $m_D = 400~\GeV$, from Eq.~\ref{eq:blind} we know $C_{h,11}^\mathrm{S}$ vanishes when $y_1 = 7.87 y_2$ or $y_1 = 0.13 y_2$. This explains a region free from SI direct detection in Fig.~\ref{fig:12cons:c}.
However, taking into account the constraints from SD direct detection and from the relic abundance, however, there is no blind spot left.

\subsubsection{LHC and LEP searches}

Searching for direct production of dark sector fermions at high energy colliders, like LHC, is another way to reveal the SDFDM model.
Due to the $Z_2$ symmetry, dark sector fermions must be produced in pairs and those other than $\chia$ eventually decay into $\chia$.
Consequently, a large missing transverse energy ($\missET$) is a typical signature for such production processes.
The $\text{monojet}+\missET$ channel could effectively probe the $\chia\chia$ pair production associated with one or two hard jets from the initial state radiation. Other dark sector pair production processes could also contribute to the $\text{monojet}+\missET$ final state if the mass spectrum is compressed.
Therefore, we should consider the following electroweak production processes for the monojet searches at the LHC:
\begin{equation}\label{eq:LHC:prod}
	pp \to \chi_i^0\chi_j^0 +\text{jets},\quad
    pp \to \chi^\pm \chi_i^0 +\text{jets},\quad
    pp \to \chi^\pm\chi^\pm +\text{jets},\quad
    i,j=1,2,3.
\end{equation}

We utilize \texttt{MadGraph 5}~\cite{Alwall:2014hca} to simulate these production processes.
\texttt{PYTHIA~6}~\cite{Sjostrand:2006za} is adopted to deal with particle decay, parton shower, and hadronization processes. \texttt{Delphes~3}~\cite{deFavereau:2013fsa} is used to carry out a fast detector simulation with a setup for the ATLAS detector.
The same cut conditions as in the ATLAS $\text{monojet}+\missET$ analysis with $20.3~\si{fb^{-1}}$ of data at $\sqrt{s}=8~\si{TeV}$~\cite{Aad:2015zva}
are applied to the above production signals in the SDFDM model.
By this way we reinterpret the experimental result to constrain the model.

In Fig.~\ref{fig:12cons}, the green regions are excluded by the $\text{monojet}+\missET$ search at 95\% CL, based on our reinterpretation.
Figs.~\ref{fig:12cons:a} and \ref{fig:12cons:b} show that the monojet search can exclude the parameter space up to $m_D\sim 80~\GeV$.
The exclusion regions hardly show dependence on $m_S$, as the singlet components in $\chi_{1,2,3}^0$ do not contribute to the production processes mediated by electroweak gauge bosons.
In Fig.~\ref{fig:12cons:c} with $m_S < m_D$, the monojet search only rules out four tiny parameter regions, because in this case $\chi_1^0$ is singlet-dominated, leading to a very low production rate for  $pp\to\chi_1^0\chi_1^0+\text{jets}$.

The charge fermion $\chi^\pm$ has similar properties as the charginos in supersymmetric models. For a rough estimation, we treat the LEP bound on the chargino mass, $m_{\tilde\chi_1^\pm}>103.5~\GeV$~\cite{Abdallah:2003xe}, as a bound on $m_{\chi^\pm}$.
As a result, the pink regions with $m_D\lesssim 100~\si{GeV}$ in Figs.~\ref{fig:12cons:a} and \ref{fig:12cons:b} are excluded.
It seems that this constraint is stronger than the monojet search at the 8~TeV LHC.

$\chi^\pm \chi_i^0$ and $\chi^\pm\chi^\pm$ production at the LHC can induce $2\ell+\missET$ and $3\ell+\missET$ signals. The leptons in the final state could be hard or soft, depending on the mass splittings $m_{\chi^\pm,\chi^0_{2,3}} - m_{\chi^0_1}$. By reinterpreting the relevant searches for hard~\cite{ATLAS:2017uun} and soft~\cite{CMS:2017fij} leptons at the 13~TeV LHC with data sets of $\sim 36~\si{fb^{-1}}$, we find that such searches cannot give stronger constraints than the bounds from the $\text{monojet}+\missET$ search and LEP.

\section{Doublet-Triplet Fermionic Dark Matter}
\label{sec:DTFDM}

In the previous section, we find that current constraints on the SDFDM model are quit severe. As a result, most of the CEPC sensitive region has already been excluded. Actually, the singlet does not have electroweak gauge interactions, so the modification to the $e^+e^-\to Zh$ cross section would not be very significant.
This observation inspires us to replace the singlet with a triplet, leading to the DTFDM model.
This model should be more capable to affect the $e^+e^-\to Zh$ cross section.
In this section, we discuss its impact on Higgs measurements at the CEPC and current constraints on its parameter space.

\subsection{Model details}
In the DTFDM model, two $\su2l$ Weyl doublets and one $\su2l$ Weyl triplet are introduced~\cite{Dedes:2014hga, Cai:2016sjz}:
\begin{equation}\label{eq:DT:rep}
	D_1 \equiv {D_1^0 \choose D_1^-} \in \left(\mathbf{2}, -\frac{1}{2}\right),\quad
    D_2 \equiv {D_2^+ \choose D_2^0} \in \left(\mathbf{2}, \frac{1}{2}\right),\quad
	T \equiv
	\left(
	\begin{array}{c}
		T^+ \\
		T^0 \\
		-T^-
	\end{array}
	\right)
	\in (\mathbf{3}, 0).
\end{equation}
We have the following gauge invariant Lagrangians:
\begin{eqnarray}
{{\cal L}_{\rm{D}}} &=& iD_1^\dag {{\bar \sigma }^\mu }{D_\mu }{D_1} + iD_2^\dag {{\bar \sigma }^\mu }{D_\mu }{D_2} - ({m_D} \epsilon_{ij} {D_1^i}{D_2^j} + \hc), \\
{{\cal L}_\mathrm{T}} &=& i{T^\dag }{{\bar \sigma }^\mu }{D_\mu }T + ({m_T}c_{ij} T^i T^j + \hc),
\end{eqnarray}
where the constants $c_{ij}$ render the gauge invariance of the $c_{ij} T^i T^j$ term.
$c_{ij}$ can be derived from Clebsch-Gordan coefficients multiplied by a factor to normalize mass terms for the components of $T$.
The nonzero values are
\begin{equation}
c_{13}=c_{31}=\frac{1}{2},\quad
c_{22}=-\frac{1}{2}.
\end{equation}

Since the hypercharge of the triplet is zero, its covariant derivative is $D_\mu=\partial_\mu - ig W_\mu^a \tau_a^{(3)}$, where $\tau_a^{(3)}$ are generators of the representation $\mathbf{3}$ for the $\mathrm{SU}(2)$ group that are chosen as
\begin{equation}
	\tau_1^{(3)} =
	\frac{1}{\sqrt{2}}\left(
		\begin{array}{ccc}
			0 & 1 &0    \\
			1 &0 & 1\\
			0 &1 &0
		 \end{array}
	\right),\quad
	\tau_2^{(3)} =
	\frac{1}{\sqrt{2}}\left(
		\begin{array}{ccc}
			0 & -i &0    \\
			i  &0 & -i\\
			0 &i &0
		 \end{array}
	\right),\quad
	\tau_3^{(3)} =
	\left(
		\begin{array}{ccc}
			1 & 0 &0    \\
			0 &0 & 0\\
			0 &0 &-1
		 \end{array}
	\right).
\end{equation}
Any irreducible $\mathrm{SU}(2)$ representation is real, in the sense that it is equivalent to its conjugate.
This equivalence means that one can find an invertible matrix $S$ satisfying $S\tau_a^{(3)}S^{-1} = - (\tau_a^{(3)})^*$.
For the generators we choose, $S$ is defined as
\begin{equation}
\begin{split}
	S =
	\left(
		\begin{array}{ccc}
			0 & 0 &-1    \\
			0 &1 & 0\\
			-1 &0 &0
		 \end{array}
	\right).
\end{split}
\end{equation}
We can use the charge conjugation matrix $\mathcal{C}=i\gamma^0\gamma^2$ to define the conjugate of the triplet as $\tilde{T}= S^{-1}\mathcal{C}\bar{T}^\mathrm{T}$, which transforms as a vector in $\mathbf{3}$, rather than in $\mathbf{\bar 3}$.
In this work, we would like to study a real triplet, which means that $\tilde{T} = T$.
This is the reason why there is a minus sign in front of the third component of $T$ in Eq.~\eqref{eq:DT:rep}.

The gauge interactions of the doublets have been explicitly listed in Eq.~\eqref{eq:2gau}, while the gauge interactions of the triplet are given by
\begin{equation}
\begin{split}
 \mathcal{L} \supset
e{A_\mu }\left[ {(T^ + )^\dag {{\bar \sigma }^\mu }{T^ + } - (T^ - )^\dag {{\bar \sigma }^\mu }{T^ - }} \right] + g c_\mathrm{W} Z_\mu \left[ {(T^ + )^\dag {{\bar \sigma }^\mu }{T^ + } - (T^ - )^\dag {{\bar \sigma }^\mu }{T^ - }} \right] \\
 + gW_\mu ^ + \left[ {(T^ + )^\dag {{\bar \sigma }^\mu }{T^0} - (T^0)^\dag {{\bar \sigma }^\mu }{T^ - }} \right] + gW_\mu ^ - \left[ {(T^0)^\dag {{\bar \sigma }^\mu }{T^ + } - (T^ - )^\dag {{\bar \sigma }^\mu }{T^0}} \right].
\end{split}
\end{equation}
The electroweak gauge symmetry allows two kinds of Yukawa couplings:
\begin{equation}
	{{\cal L}_{{\rm{Y}}}} = {y_1} c_{ijk} T^i D_1^j H^k - {y_2} c_{ijk} T^i D_2^j  H^k + \hc ,
\end{equation}
where the constants $c_{ijk}$ can also be built from Clebsch-Gordan coefficients. Their nonzero values are
\begin{equation}
c_{122}=c_{311}=\sqrt{2},\quad
c_{212}=c_{221}=-1.
\end{equation}

\begin{figure}[!tbp]
\centering
\subfigure[~$m_T = 100~\GeV$, $m_D = 400~\GeV$.]
{\includegraphics[width=.45\textwidth]{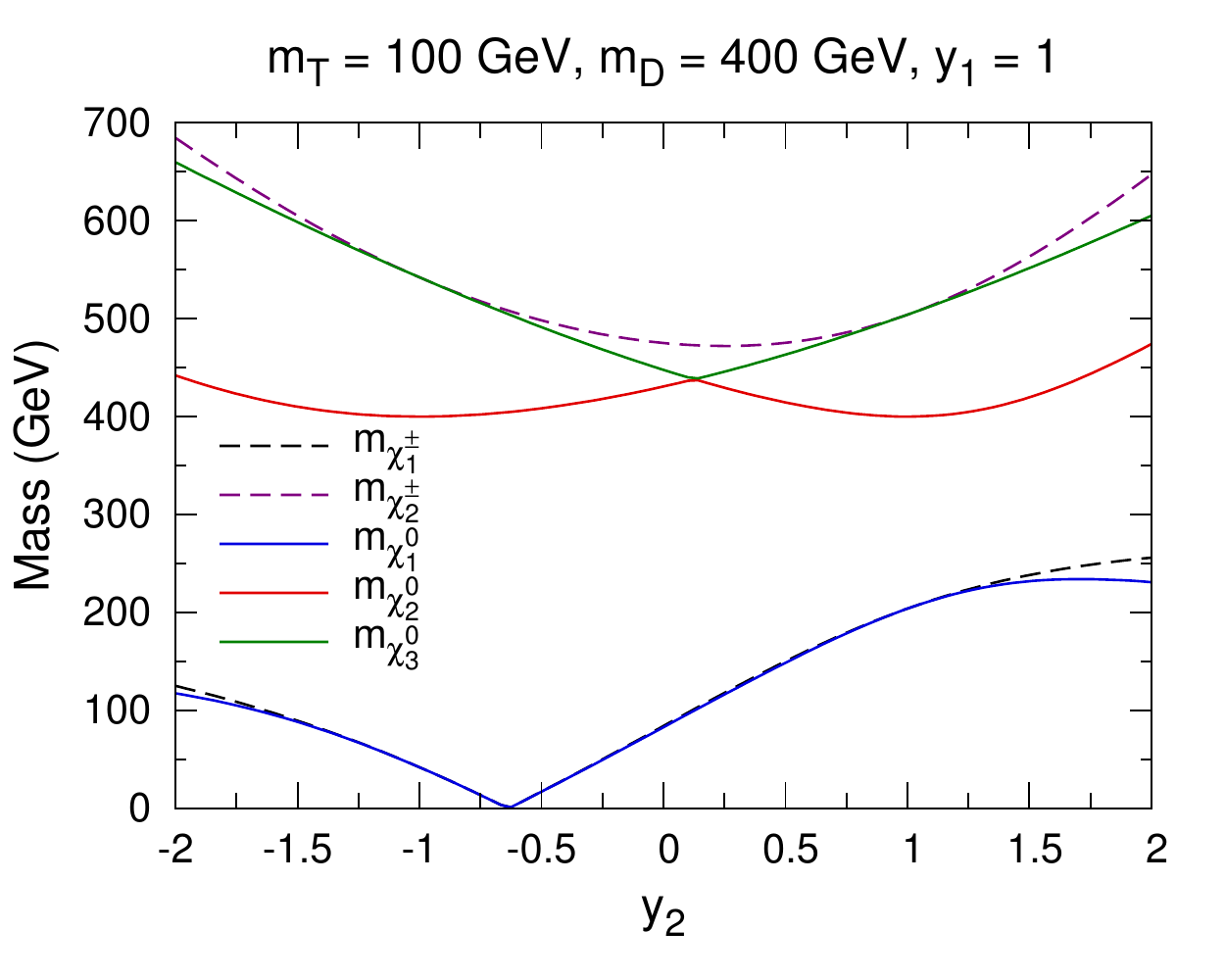}}
\subfigure[~$m_T = 400~\GeV$, $m_D = 150~\GeV$.]
{\includegraphics[width=.45\textwidth]{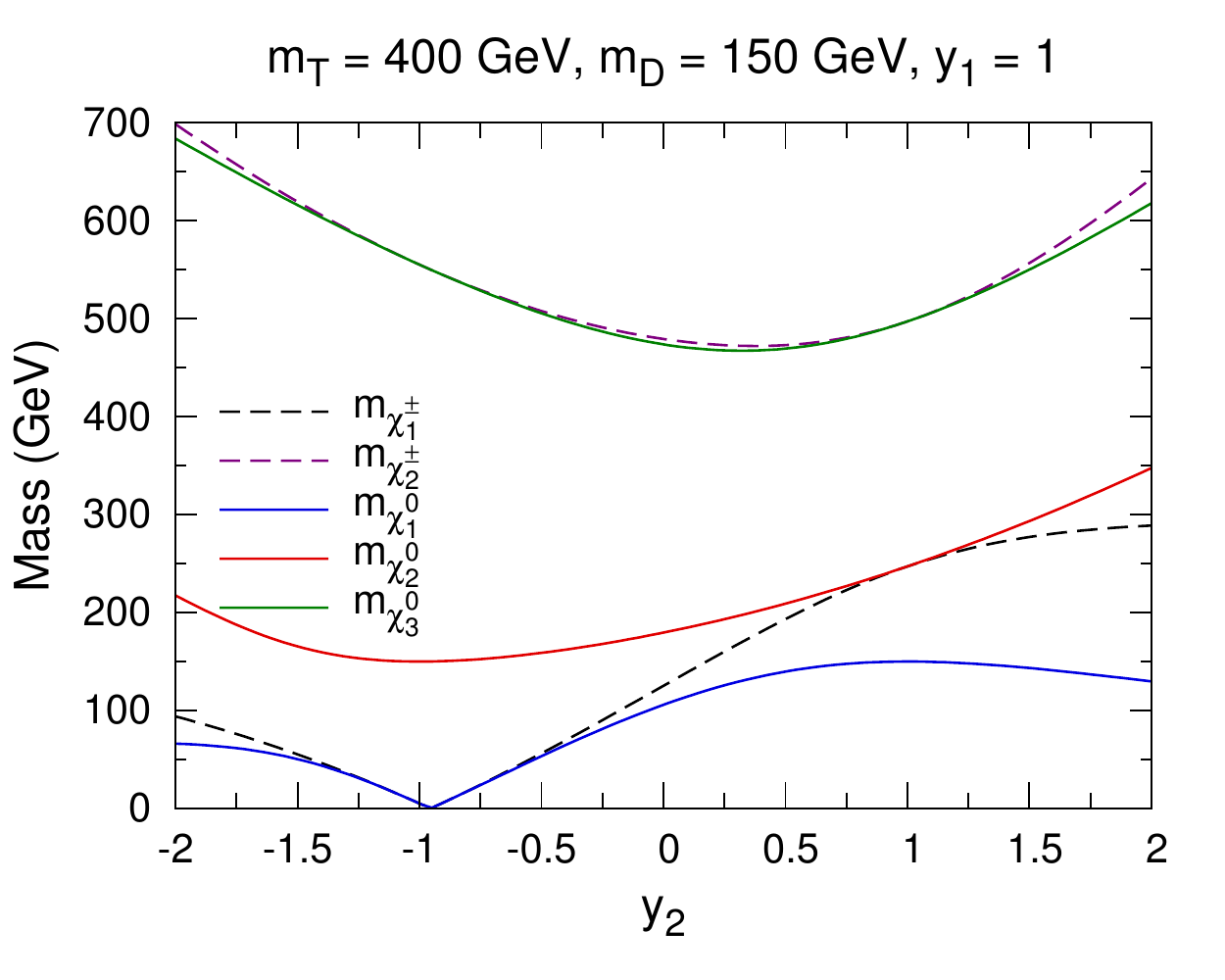}}
\caption{Mass spectra of the DTFDM model in two typical cases, $m_T < m_D$ (a) and $m_T > m_D$ (b).}
\label{fig:23spec}
\end{figure}

After the Higgs field develops a VEV, mass terms in the dark sector can be expressed as
\begin{equation}
	\begin{split}
		\mathcal{L}_\mathrm{M} =
		& -\frac{1}{2}
		\left(
			\begin{array}{ccc}
				T & D_1^0 & D_2^0
			\end{array}
		\right)
		\mathcal{M}_\mathrm{N}
		\left(
			\begin{array}{c}
				T \\ D_1^0 \\ D_2^0
			\end{array}
		\right)
	 -  \left(
	 		\begin{array}{cc}
				T^- & D_1^-
			\end{array}
	   \right)
		\mC
		\left(
			\begin{array}{c}
				T^+ \\ D_2^+
			\end{array}
		\right)
	+ \hc \\
	= & - \frac{1}{2} \sum_{i=1}^3 m_{\chiaa} \chiaa \chiaa - \sum_{i=1}^2 m_{\chii} \chi_i^- \chi_i^+ + \hc
	\end{split}
\end{equation}
The mass and mixing matrices are defined as
\begin{eqnarray}
	\mN &=&
	\left(
		\begin{array}{ccc}
			m_T                      & \dfrac{1}{\sqrt{2}} y_1 v & \dfrac{1}{\sqrt{2}} y_2 v \\[1em]
			\dfrac{1}{\sqrt{2}} y_1 v & 0                        & -m_D \\[1em]
			 \dfrac{1}{\sqrt{2}} y_2 v & -m_D                    &0
		 \end{array}
	\right),\quad
	\mC
	\left(
	\begin{array}{cc}
		m_T    ~~  & y_2 v \\
		-y_1 v ~~  & m_D
	\end{array}
	\right).
\\
	\N^\mathrm{T} \mN \N &=& \mathrm{diag}(m_{\chia}, m_{\chib}, m_{\chic}), \quad
	\CR^\mathrm{T} \mC \CL = \mathrm{diag} (m_{\chiapm}, m_{\chibpm}).
\\
	\left(
	\begin{array}{c}
		T^0 \\ D_1^0 \\ D_2^0
	\end{array}
	\right)
	&=&
	\N
	\left(
	\begin{array}{c}
		\chia \\ \chib \\ \chic
	\end{array}
	\right),\quad
	\left(
	    \begin{array}{c} T^+ \\ D_2^+ \end{array}
	\right)
		 = \CL
	\left(
	    \begin{array}{c} \chi_1^+ \\ \chi_2^+ \end{array}
	\right),\quad
	\left(
	    \begin{array}{c} T^- \\ D_1^- \end{array}
	\right)
		 = \CR
	\left(
	    \begin{array}{c} \chi_1^- \\ \chi_2^- \end{array}
		\right).
\end{eqnarray}
Thus, the dark sector contains three Majorana fermions $\chi_{1,2,3}^0$ and two charged Dirac fermions $\chi_{1,2}^\pm$.
In Fig.~\ref{fig:23spec} we show the mass spectra for two typical cases, $m_T < m_D$ and $m_T > m_D$.
The masses of neutral fermions have the similar behavior as in the SDFDM model, since $\mathcal{M}_\mathrm{N}$ is the same if $m_T$ is replaced by $m_S$. Nonetheless, the masses of charged fermions vary with $y_2$ due to the mixing, unlike $\chi^\pm$ in the SDFDM model.

By defining Dirac spinors $\Psi_{1,2}^+$ and Majorana spinors $\Psi_{1,2,3}$ as
\begin{equation}
\Psi_i^+ = \begin{pmatrix}
\chi_i^+\\
(\chi_i^-)^\dag
\end{pmatrix},\quad
\Psi_i = \begin{pmatrix}
\chi_i^0\\
(\chi_i^0)^\dag
\end{pmatrix},
\end{equation}
we have the following interaction terms:
\begin{eqnarray}
		\mathcal{L}_\mathrm{int} &=& e \sum_{i} A_\mu \bar{\Psi}_i^+ \gamma^\mu \Psi_i^+
		+  \sum_{ij} Z_\mu ( G_{Z,ij}^\mathrm{L} \bar{\Psi}_i^+ \gamma^\mu P_\mathrm{L} \Psi_j^+
		                               +  G_{Z,ij}^\mathrm{R} \bar{\Psi}_i^+ \gamma^\mu P_\mathrm{R} \Psi_j^+ )\nonumber\\
	&& +\sum_{ij} G_{h,ij}^\mathrm{S} h \bar{\Psi}_i^+ \Psi_j^+
	  -\sum_{ij} G_{h,ij}^\mathrm{P} h \bar{\Psi}_i^+ i \gamma^5 \Psi_j^+ \nonumber\\
	&& + \sum_{ij} W_\mu^- (G_{W,ij}^\mathrm{L} \bar{\Psi}_i \gamma^\mu P_\mathrm{L} \Psi_j^+
	                                     - G_{W,ij}^\mathrm{R} \bar{\Psi}_i \gamma^\mu P_\mathrm{R} \Psi_j^+) \nonumber\\
	&& + \sum_{ij} W_\mu^+ (G_{W,ij}^{\mathrm{L}\ast} \bar{\Psi}_j^+ \gamma^\mu P_\mathrm{L} \Psi_i
	    - G_{W,ij}^{\mathrm{R}\ast} \bar{\Psi}_j^+ \gamma^\mu P_\mathrm{R} \Psi_i) \nonumber\\
	&& -\frac{1}{2} \sum_{ij} C_{Z,ij}^\mathrm{A} Z_\mu \bar{\Psi}_i \gamma^\mu \gamma^5 \Psi_j
	  + \frac{1}{2} \sum_{ij} C_{Z,ij}^\mathrm{V} Z_\mu \bar{\Psi}_i  \gamma^\mu \Psi_j \nonumber\\
	&& -\frac{1}{2} \sum_{ij} C_{h,ij}^\mathrm{S} h \bar{\Psi}_i \Psi_j
	  +\frac{1}{2} \sum_{ij} C_{h,ij}^\mathrm{P} h \bar{\Psi}_i i \gamma^5 \Psi_j .
\end{eqnarray}
The couplings are defined as
\begin{eqnarray}
	G_{Z,ij}^\mathrm{L} &=& \frac{g(c_\mathrm{W}^2-s_\mathrm{W}^2)}{ 2 c_\mathrm{W}}\mathcal{C}_{\mathrm{L},2i}^\ast \mathcal{C}_{\mathrm{L},2j}
	+ g c_\mathrm{W} \mathcal{C}_{\mathrm{L},1i}^\ast \mathcal{C}_{\mathrm{L},1j},\\
	G_{Z,ij}^\mathrm{R} &=&
    \frac{g(c_\mathrm{W}^2-s_\mathrm{W}^2)}{ 2 c_\mathrm{W}}\mathcal{C}_{\mathrm{R},2j}^\ast \mathcal{C}_{\mathrm{R},2i}
	+ g c_\mathrm{W} \mathcal{C}_{\mathrm{R},1j}^\ast \mathcal{C}_{\mathrm{R},1i},\\
	G_{h,ij}^\mathrm{S} &=& \mathrm{Re}(y_1 \mathcal{C}_{\mathrm{L},1j} \mathcal{C}_{\mathrm{R},2i} - y_2 \mathcal{C}_{\mathrm{L},2j} \mathcal{C}_{\mathrm{R},1i}),
	\quad G_{h,ij}^\mathrm{P} = \mathrm{Im}(y_1 \mathcal{C}_{\mathrm{L},1j} \mathcal{C}_{\mathrm{R},2i} - y_2 \mathcal{C}_{\mathrm{L},2j} \mathcal{C}_{\mathrm{R},1i}),\\
	G_{W,ij}^\mathrm{L} &=&   \frac{g}{\sqrt{2}}\N_{3i}^\ast \mathcal{C}_{\mathrm{L},2j} + g \N_{1i}^\ast \mathcal{C}_{\mathrm{L},1j},
	\quad
    G_{W,ij}^\mathrm{R} =   \frac{g}{\sqrt{2}}\N_{2i} \mathcal{C}_{\mathrm{R},2j}^\ast - g \N_{1i} \mathcal{C}_{\mathrm{R},1j}^\ast, \\
	C_{Z,ij}^\mathrm{A} &=& \frac{g}{2 c_\mathrm{W}} \mathrm{Re}(\N_{2i}^\ast \N_{2j} - \N_{3i}^\ast \N_{3j}), \quad
	C_{Z,ij}^\mathrm{V} = \frac{ig}{2 c_\mathrm{W}} \mathrm{Im}(\N_{2i}^\ast \N_{2j} - \N_{3i}^\ast \N_{3j}),\\
	C_{h,ij}^\mathrm{S} &=& \sqrt{2}~\mathrm{Re}(y_1 \N_{1i} \N_{2j} + y_2 \N_{1i} \N_{3j}), \quad
	C_{h,ij}^\mathrm{P} = \sqrt{2}~\mathrm{Im}(y_1 \N_{1i} \N_{2j} + y_2 \N_{1i} \N_{3j}).
\end{eqnarray}
Note that $C_{Z,ij}^\mathrm{A}$, $C_{Z,ij}^\mathrm{V}$, $C_{h,ij}^\mathrm{S}$, and $C_{h,ij}^\mathrm{P}$ have the same forms as those in the SDFDM model, because $T^0$ has neither electrical charge nor hypercharge, just like the singlet $S$.
Consequently, $y_1=\pm y_2$ also leads to $C_{Z,11}^\mathrm{A}=0$, while $y_1=y_2$ and $m_D<m_T$ lead to $C_{h,11}^\mathrm{S}=0$.
Thus, the sensitivity of DM direct detection to this model should be similar to the SDFDM model.

\subsection{Higgs Precision Measurements at the CEPC}

\subsubsection{Corrections to $Zh$ associated production}

\begin{figure}[!tbp]
\centering
\subfigure[\label{fig:DT:v}]
{\includegraphics[width=.3\textwidth]{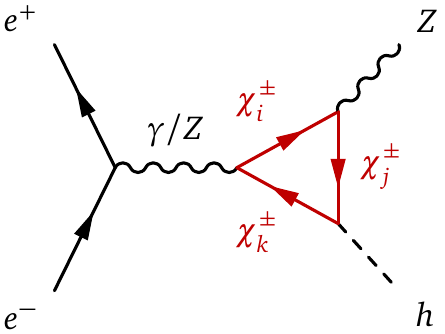}}
\hspace{3em}
\subfigure[\label{fig:DT:hh}]
{\includegraphics[width=.28\textwidth]{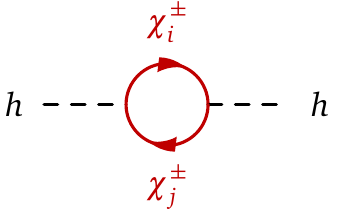}}
\caption{Feynman diagrams for vertex (a) and self-energy (b) corrections to $e^+ e^- \to Z h$ at one-loop level due to the $h\chi_i^\pm\chi_j^\pm$ couplings in the DTFDM model.}
\label{fig:DT:fd}
\end{figure}

In the DTFDM model, the $e^+e^-\to Zh$ process is modified at one-loop level by the Feynman diagrams shown in Figs.~\ref{fig:v} and \ref{fig:Z} with $\chi^\pm$ replaced by $\chi^\pm_{1,2}$.
Unlike the SDFDM model, however, the charged dark sector fermions in the DTFDM model can couple to the Higgs boson, because both $D$ and $T$ involve charged components.
Consequently, we also have the vertex corrections shown in Fig.~\ref{fig:DT:v} and the self-energy corrections shown in Fig.~\ref{fig:DT:hh}.
Because more dark sector fermions could influence the $Zh$ associated production, a larger modification of the $e^+e^-\to Zh$ cross section is expected.

\begin{figure}[!tbp]
\centering
\subfigure[~$y_1 = 0.5$, $y_2 = 1.5$.\label{fig:23zh:a}]
{\includegraphics[width=.45\textwidth,trim={0 30 0 10},clip]{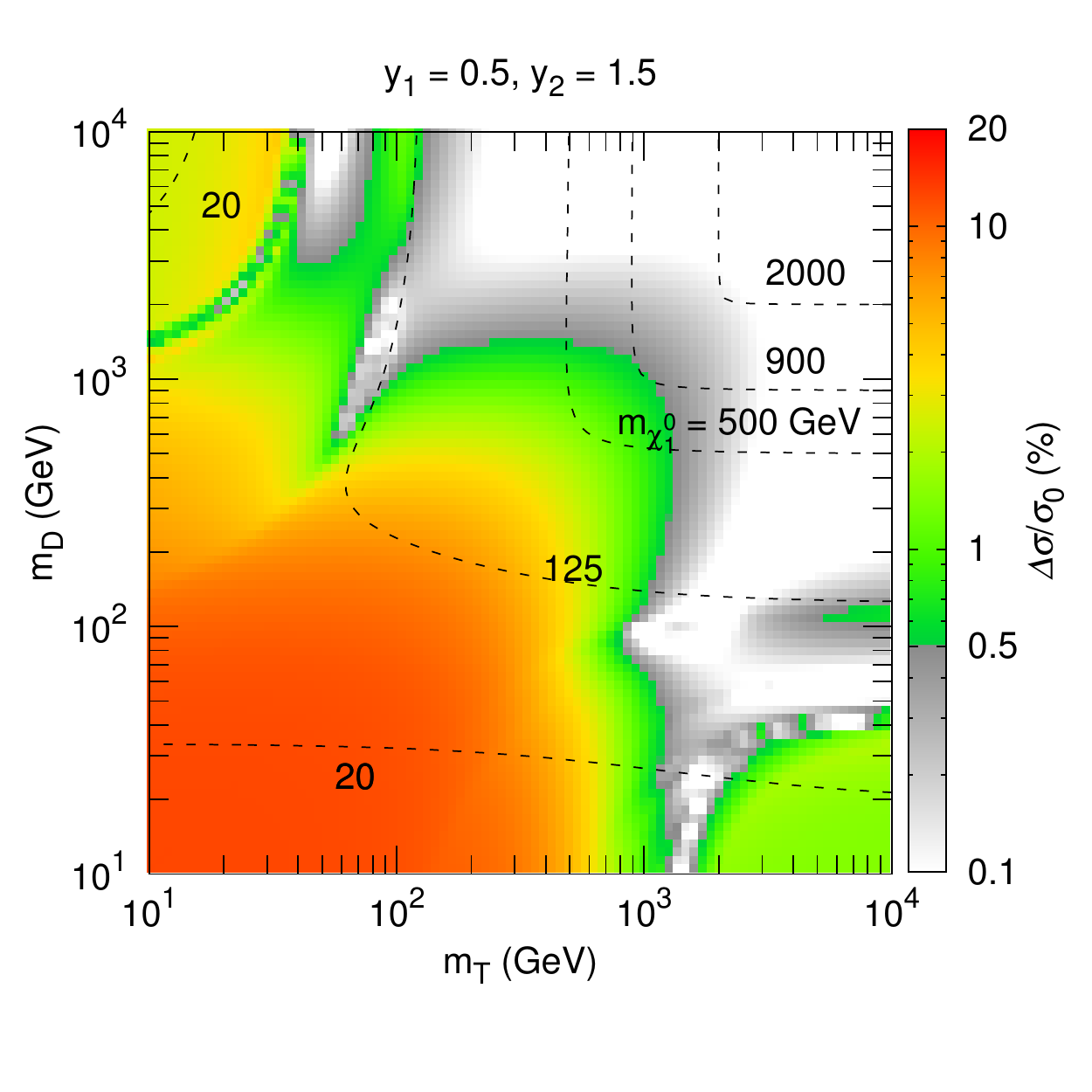}}
\subfigure[~$y_1 = 1.0$, $y_2 = 1.0$.\label{fig:23zh:b}]
{\includegraphics[width=.45\textwidth,trim={0 30 0 10},clip]{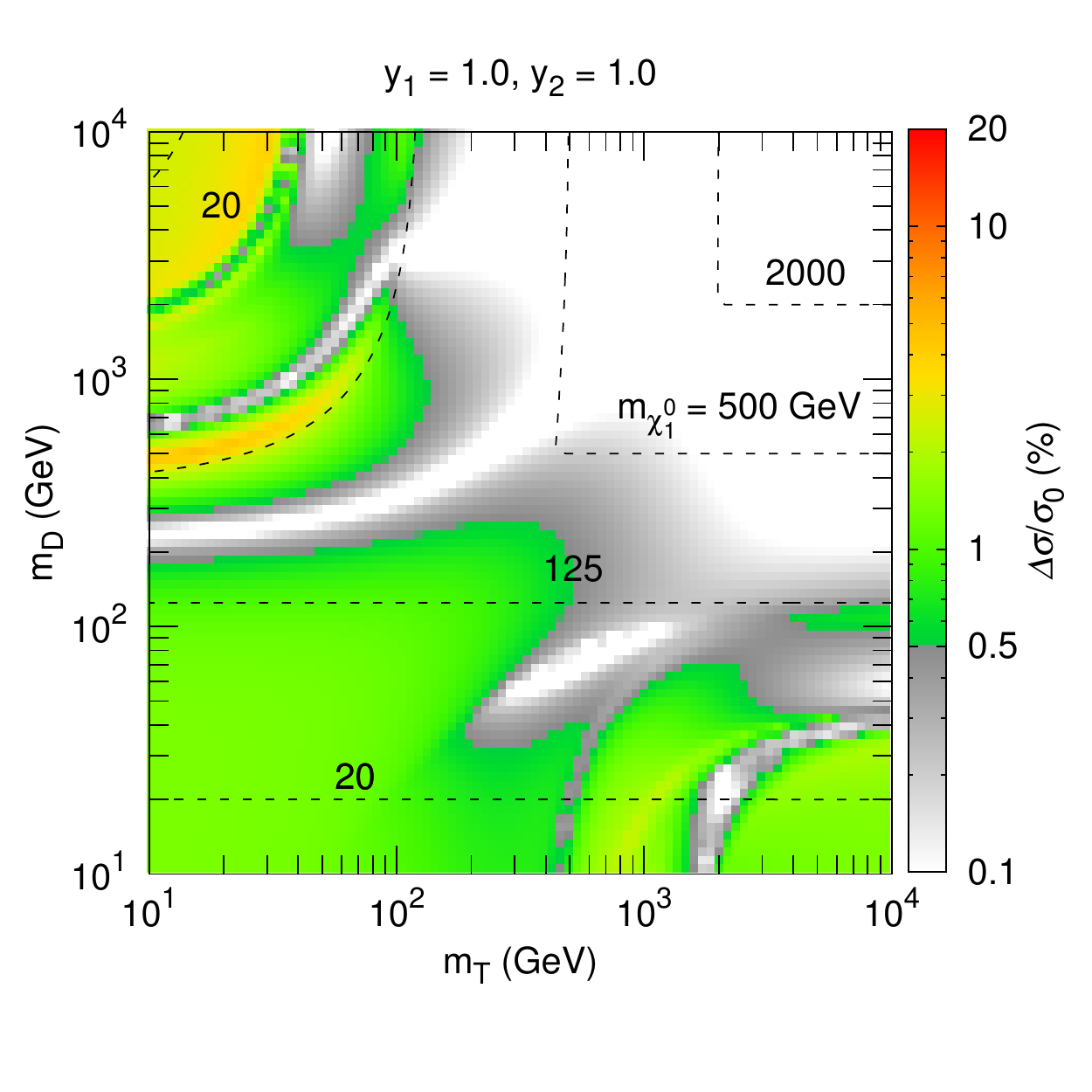}}
\subfigure[~$m_T = 100~\GeV$, $m_D = 400~\GeV$.\label{fig:23zh:c}]
{\includegraphics[width=.45\textwidth,trim={0 30 0 10},clip]{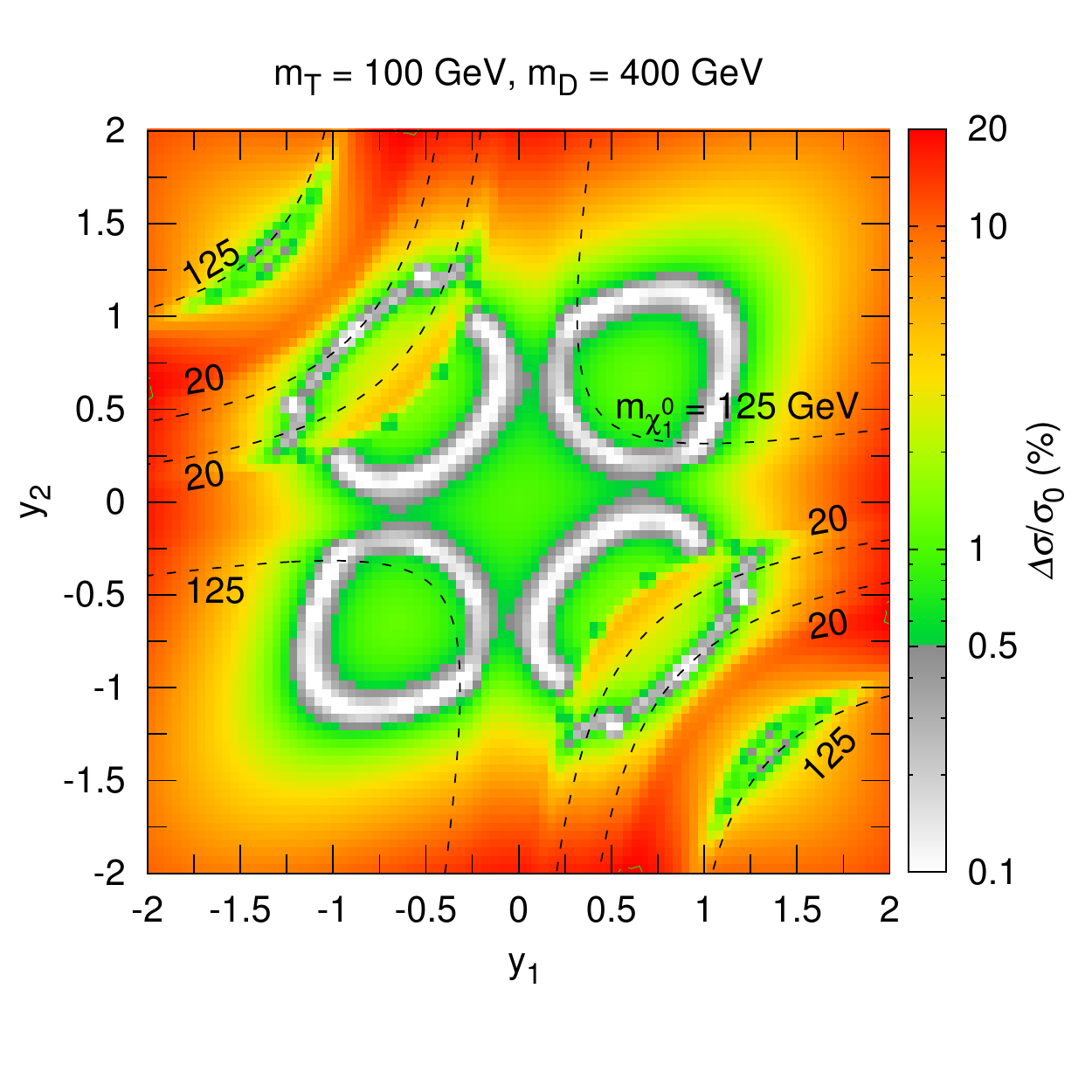}}
\subfigure[~$m_T = 400~\GeV$, $m_D = 150~\GeV$.\label{fig:23zh:d}]
{\includegraphics[width=.45\textwidth,trim={0 30 0 10},clip]{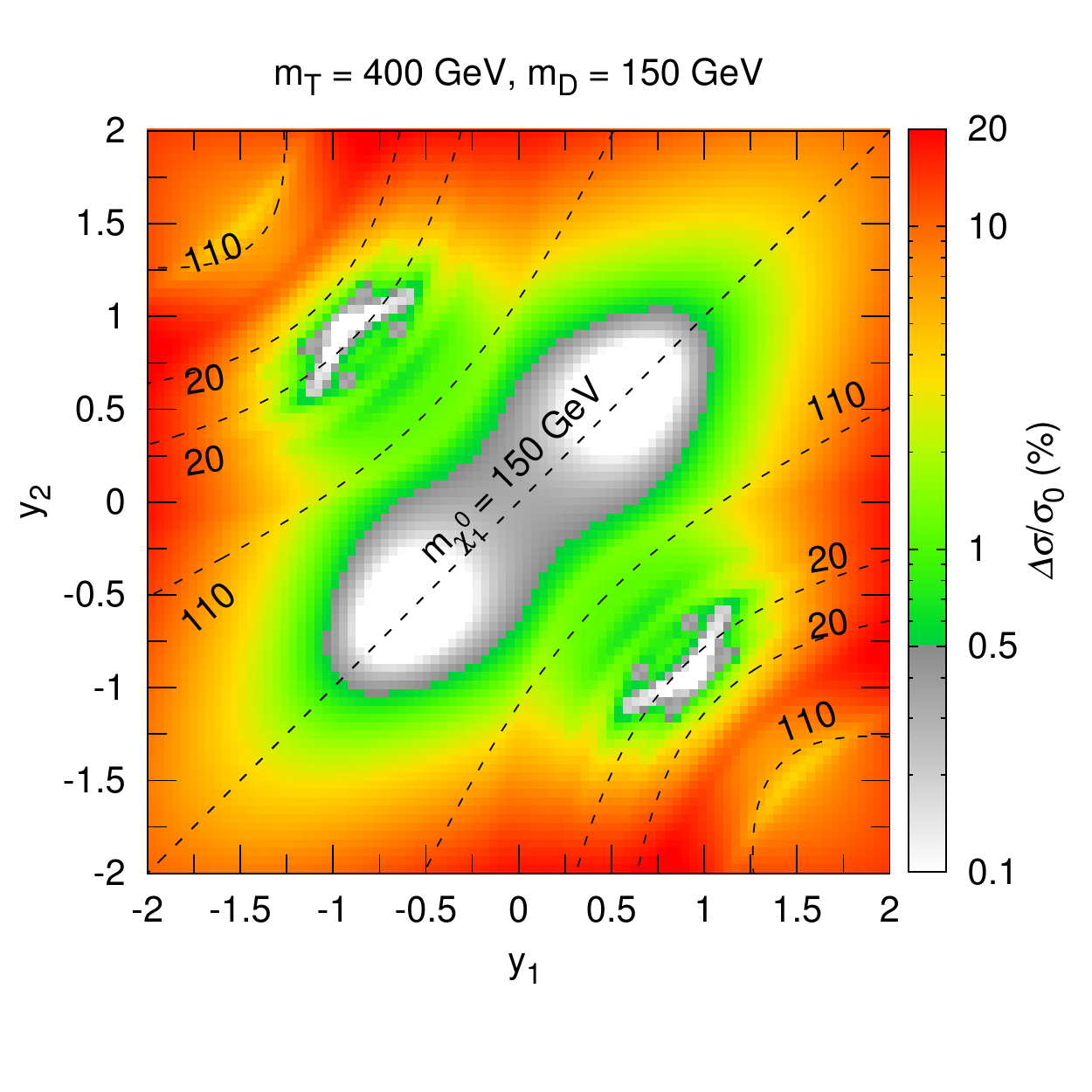}}
\caption{Heat maps for $\Delta \sigma/\sigma_0$ in the DTFDM model. Results are shown in the $m_T-m_D$ (a,b) and $y_1-y_2$ (c,d) planes with two parameters fixed as indicated.
Colored and gray regions correspond to $\Delta \sigma/\sigma_0 > 0.5\%$ and $< 0.5\%$, respectively.
Dashes lines denote contours of the DM candidate mass $\mchia$.}
\label{fig:23zh}
\end{figure}

In Fig.~\ref{fig:23zh}, we show the absolute relative deviation of the $e^+e^-\to Zh$ cross section $\Delta \sigma/\sigma_0$ in the DTFDM model.
Compared with Fig.~\ref{fig:12zh} in the SDFDM model, the deviation generally increases.
As illustrated in Figs.~\ref{fig:23zh:a} and \ref{fig:23zh:b}, the deviation in the regions with a small $m_T$ and a large $m_D$ can be significant. In contrary, we should recall that a small $m_S$ and a large $m_D$ would lead to an unreachable deviation shown in Figs.~\ref{fig:12zh:a} and \ref{fig:12zh:b}.
This clearly demonstrates the effect of the substitution of the triplet for the singlet.
Fig.~\ref{fig:23zh:a} indicates that the CEPC measurement of $e^+e^-\to Zh$ could explore up to $m_{\chia}\sim 900~\GeV$ for $y_1 = 0.5$ and $y_2 = 1.5$.
Moreover, there are only a few small regions with $\Delta \sigma/\sigma_0<0.5\%$ in Figs.~\ref{fig:23zh:c} and \ref{fig:23zh:d}.

\subsubsection{Higgs boson invisible decay}

In the DTFDM model, the $h$ and $Z$ decay widths into $\chi_i^0\chi_j^0$  ($i,j=1,2,3$) have the same expressions as Eqs.~\eqref{eq:h2chi0ichi0j}--\eqref{eq:Z2chi0ichi0i}, while the $Z$ decay widths into $Z\to\chi_i^+\chi_j^-$ ($i,j=1,2$) are given by
\begin{eqnarray}
\Gamma (Z\to\chi_i^+\chi_j^-) &=& \frac{{F(m_Z^2,m_{\chi _i^ \pm }^2,m_{\chi _j^ \pm }^2)}}{{48\pi m_Z^5}}\big\{ 6(G_{Z,ij}^{\mathrm{L}}G_{Z,ij}^{{\mathrm{R*}}} + G_{Z,ij}^{{\mathrm{L*}}}G_{Z,ij}^{\mathrm{R}})m_Z^2{m_{\chi _i^ \pm }}{m_{\chi _j^ \pm }} \nonumber\\
&& + (|G_{Z,ij}^{\mathrm{L}}{|^2} + |G_{Z,ij}^{\mathrm{R}}{|^2})[m_Z^2(2m_Z^2 - m_{\chi _i^ \pm }^2 - m_{\chi _j^ \pm }^2) - {(m_{\chi _i^ \pm }^2 - m_{\chi _j^ \pm }^2)^2}]\big\}.
\end{eqnarray}
Furthermore, the $h\chi_i^\pm\chi_j^\pm$ couplings could induce Higgs boson decay channels into $\chi_i^+\chi_j^-$ if the kinematics is allowed. The corresponding widths are
\begin{eqnarray}
\Gamma (h\to\chi_i^+\chi_j^-) &=&
\frac{{F(m_h^2,m_{\chi _i^ \pm }^2,m_{\chi _j^ \pm }^2)}}{{8\pi m_h^3}}\big\{ |G_{h,ij}^{\mathrm{S}}{|^2}[m_h^2 - {({m_{\chi _i^ \pm }} + {m_{\chi _j^ \pm }})^2}] \nonumber\\
&&+ |G_{h,ij}^{\mathrm{P}}{|^2}[m_h^2 - {({m_{\chi _i^ \pm }} - {m_{\chi _j^ \pm }})^2}]\big\}.
\end{eqnarray}

\begin{figure}[!tbp]
\centering
\subfigure[~$y_1 = 0.5$, $y_2 = 1.5$.]
{\includegraphics[width=.45\textwidth,trim={0 15 0 10},clip]{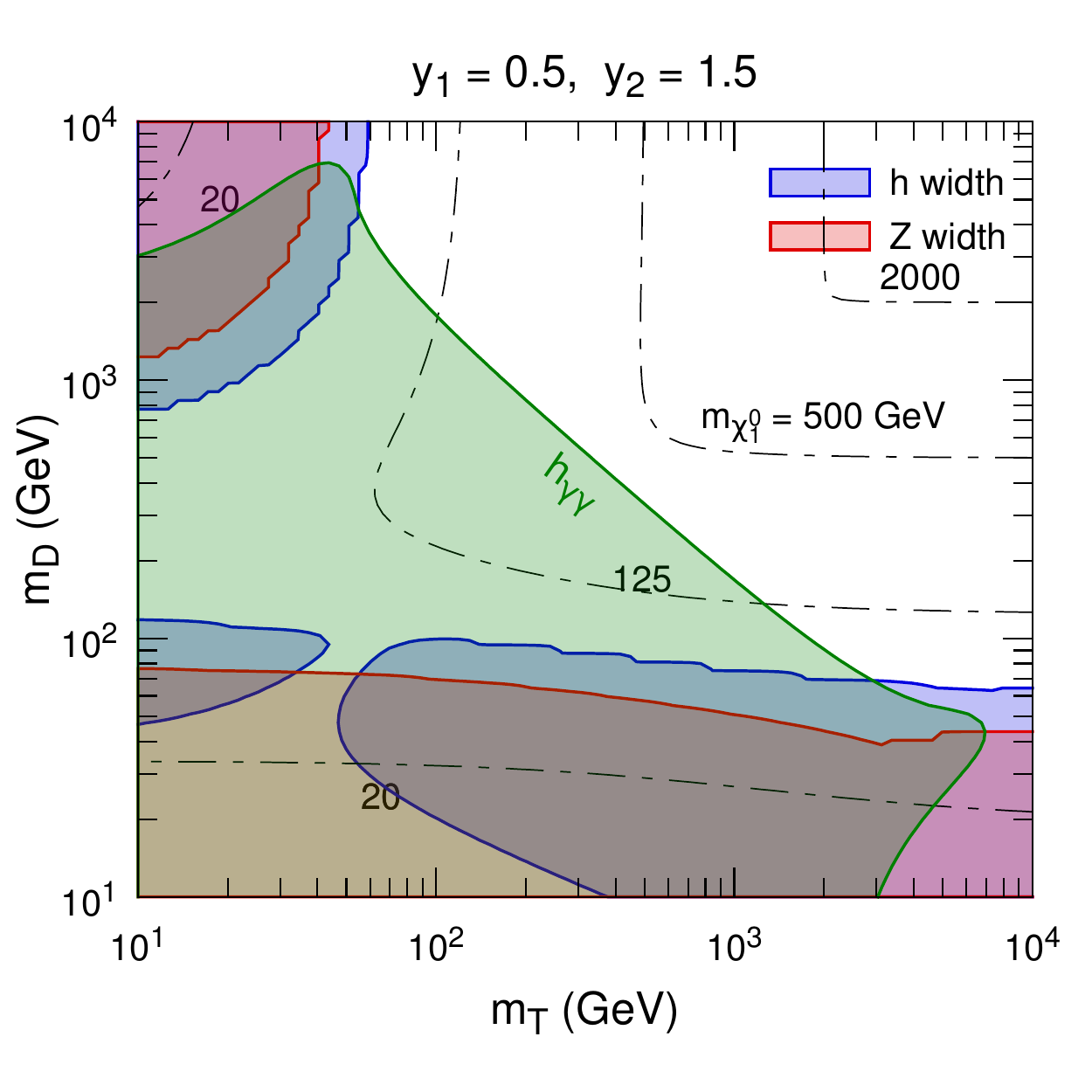}}
\subfigure[~$y_1 = y_2 = 1.0$.]
{\includegraphics[width=.45\textwidth,trim={0 15 0 10},clip]{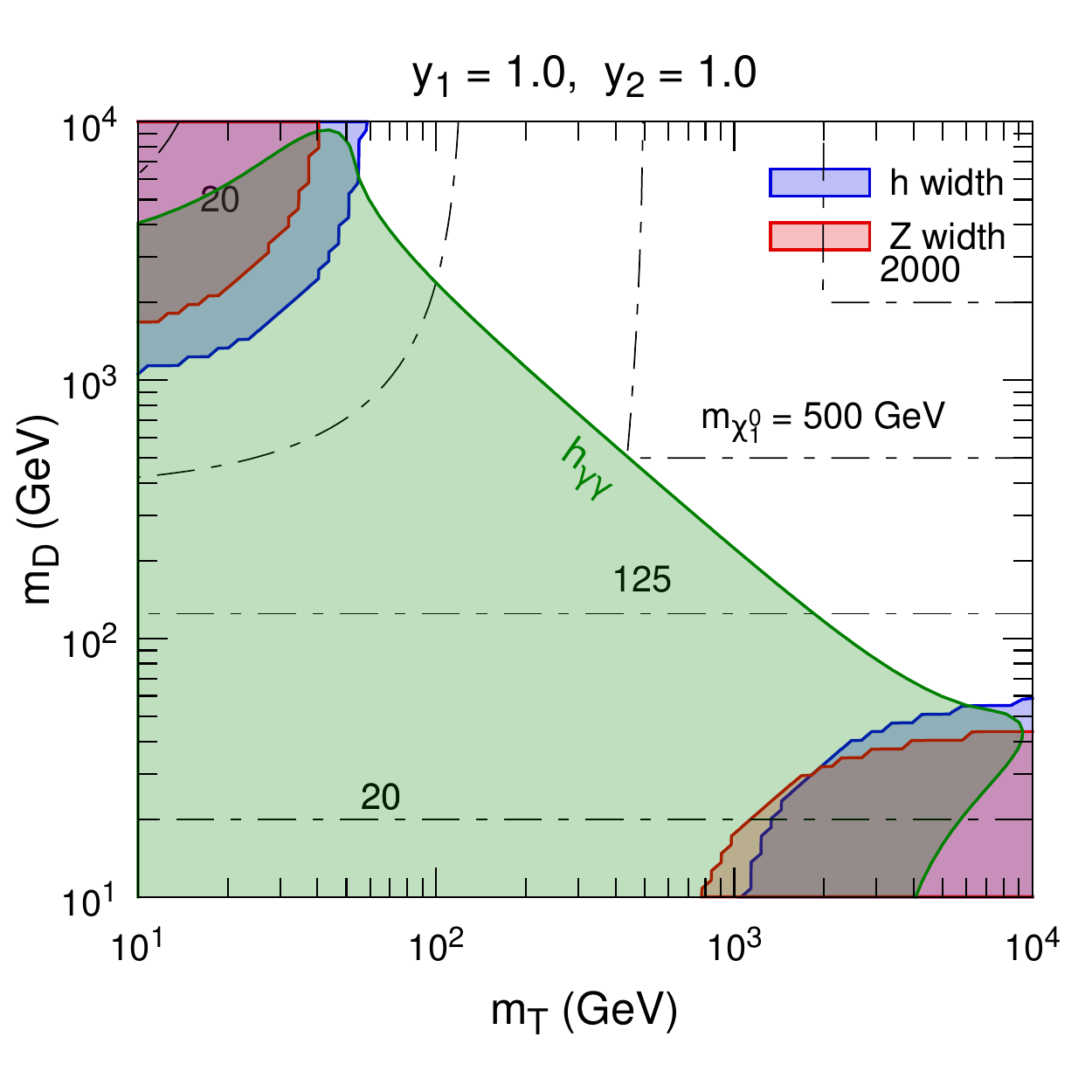}}
\subfigure[~$m_T = 100~\GeV$, $m_D = 400~\GeV$.]
{\includegraphics[width=.45\textwidth,trim={0 15 0 10},clip]{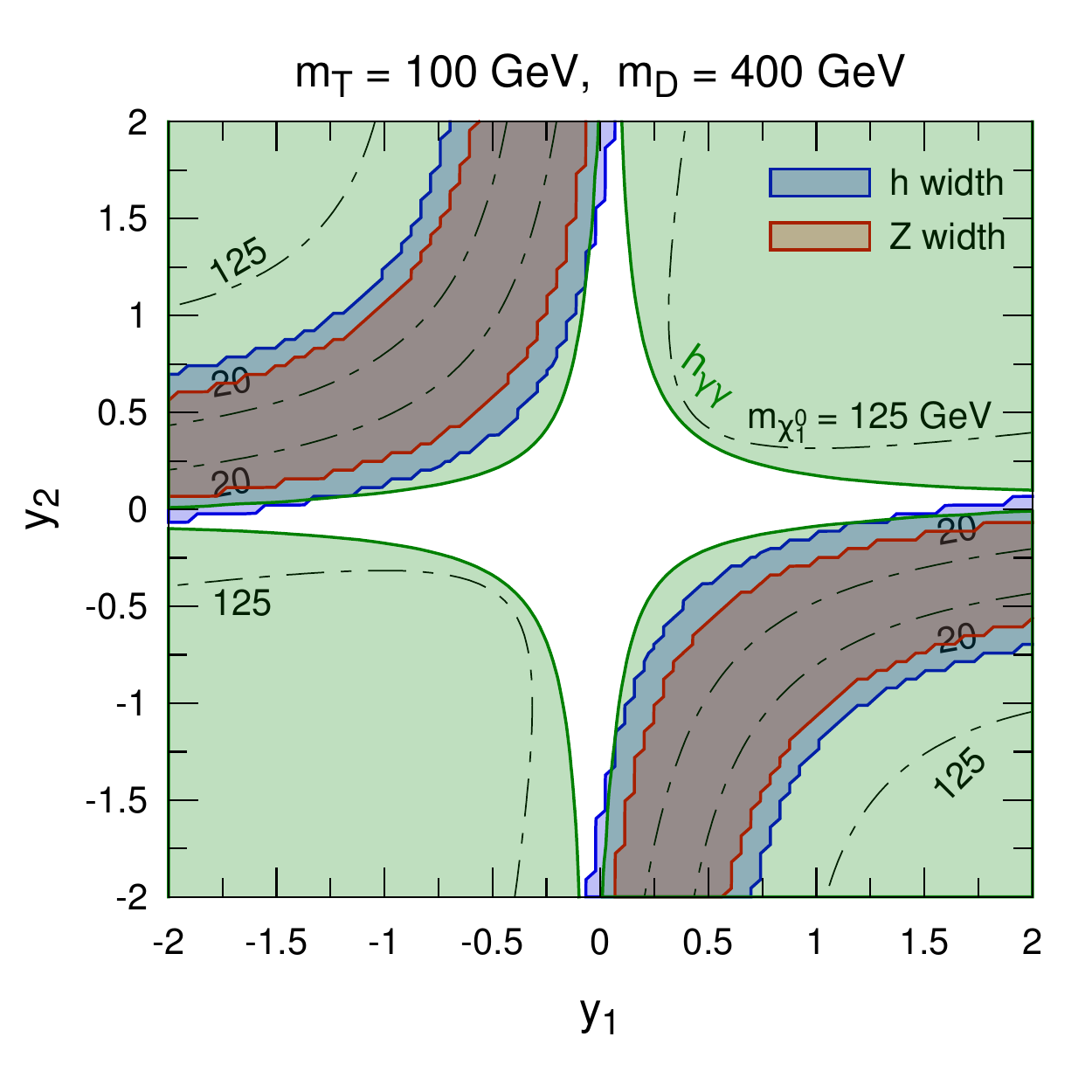}}
\subfigure[~$m_T = 400~\GeV$, $m_D = 150~\GeV$.]
{\includegraphics[width=.45\textwidth,trim={0 15 0 10},clip]{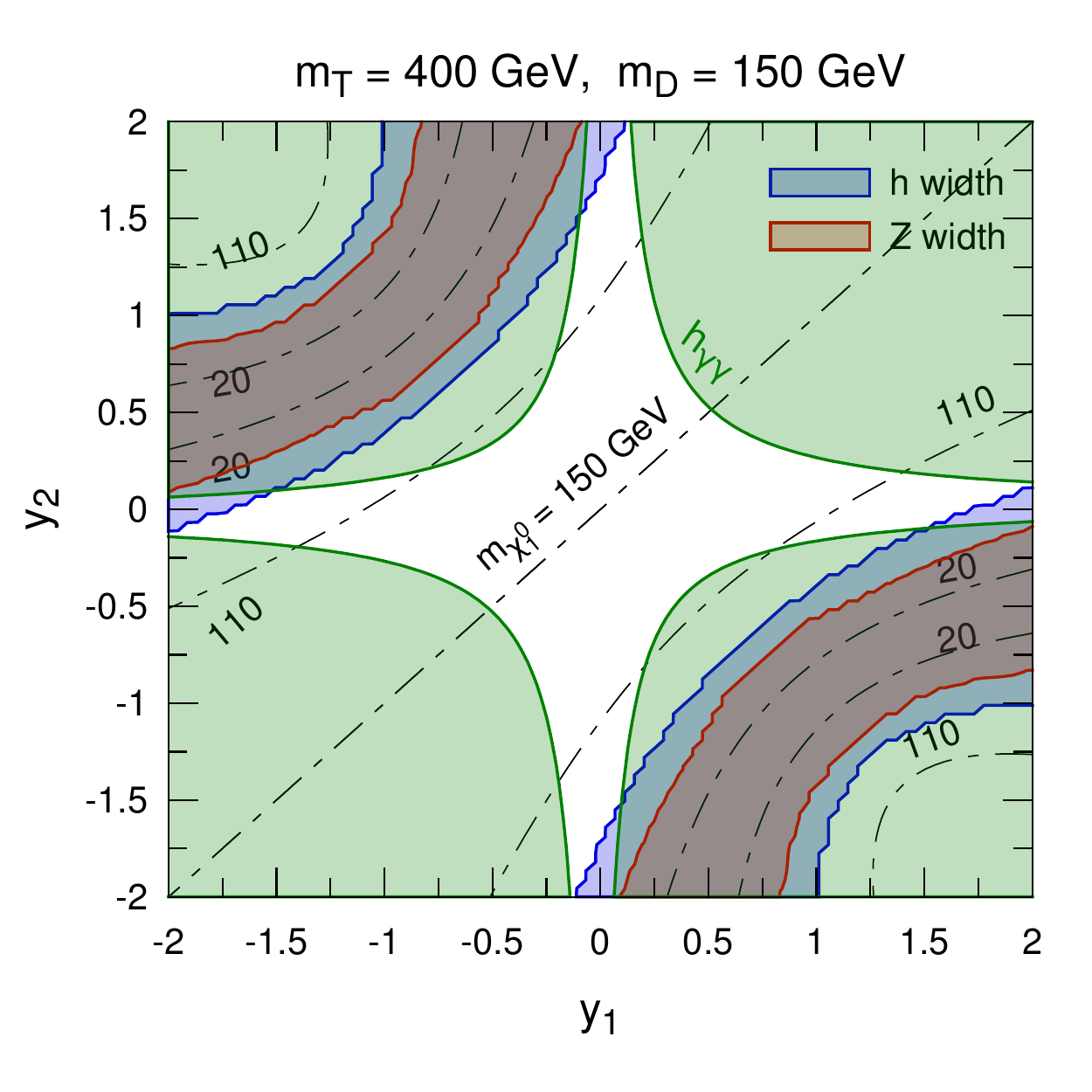}}
\caption{95\% CL expected constraints from the CEPC measurements of the Higgs boson invisible (blue regions) and diphoton (green regions) decay widths
in the $m_T-m_D$ plane (a,b) and the $y_1-y_2$ plane (c,d) for the DTFDM model.
Red regions have been excluded at 95\% CL by the LEP measurement of the $Z$ boson invisible decay width~\cite{ALEPH:2005ab}.
Dot-dashed lines indicate $m_{\chi_1^0}$ contours.}
\label{fig:23width}
\end{figure}

In Fig.~\ref{fig:23width} we present the expected CEPC constraint from the $h$ invisible decay as well as the LEP constraint from the $Z$ invisible decay.
Compared with Fig.~\ref{fig:12width} for the SDFDM model, the LEP exclusion regions for the DTFDM model are enlarged because of more $Z$ decay channels.
On the other hand, the CEPC sensitivities are quit similar in both models.

\subsubsection{Higgs boson diphoton decay}
\label{sec:haa}

\begin{figure}[!tbp]
\centering
\includegraphics[width=.28\textwidth]{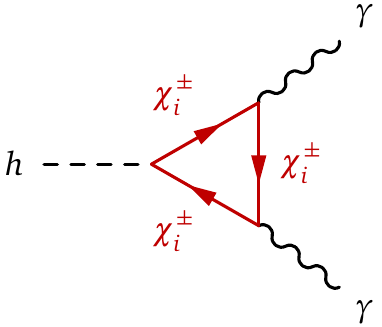}
\caption{Feynman diagram for $h\to\gamma\gamma$ due to $\chi^\pm_i$ loops in the DTFDM model.}
\label{fig:DT:fd:h2aa}
\end{figure}

Another remarkable feature of the DTFDM model is that the $h\chi_i^\pm\chi_i^\pm$ and $\gamma\chi_i^\pm\chi_i^\pm$ couplings modify the width of the Higgs boson diphoton decay, $h\to\gamma\gamma$, at one-loop level.
Fig.~\ref{fig:DT:fd:h2aa} demonstrates the related Feynman diagram.
As CEPC can accurately measure the relative precision of the $h\to\gamma\gamma$ decay width down to $9.4\%$\footnote{This is a conservative value; if one considers a combination with the high-luminosity LHC measurement, the relative precision can be improved to 4.6\%~\cite{CEPC-SPPCStudyGroup:2015csa}.} with an integrated luminosity of $5~\iab$~\cite{CEPC-SPPCStudyGroup:2015csa}, this decay channel could be very sensitive to the DTFDM model.

At the leading order in the SM, the  Higgs boson decay into two photons is induced by loops, mediated by the $W$ boson and heavy charged fermions.
In the DTFDM model, we should also take into account the $\chiapm$ and $\chibpm$ loops.
Thus, the $h\to\gamma\gamma$ partial decay width can be expressed as~\cite{Ellis:1975ap,Shifman:1979eb}
\begin{equation}
	\Gamma (h \to \gamma \gamma) = \frac{G_\mathrm{F} \alpha^2 m_h^3}{128\sqrt{2}\pi^3}
	\left| A_1 (\tau_W) + \sum_f c_f Q_f^2 A_{1/2} (\tau_f)
     + \sum_i \frac{G^\mathrm{S}_{h,ii}v}{m_{\chii}} A_{1/2}(\tau_{\chii}) \right|^2,
\end{equation}
where $G_\mathrm{F}$ is the Fermi coupling constant and $\alpha$ is the fine-structure constant. $c_f$ and $Q_f$ are the color factor and the electric charge of an SM fermion $f$, respectively
The form factors $A_1(\tau)$ and $A_{1/2}(\tau)$ are defined as
\begin{equation}
	A_1 (\tau) = - \tau^{-2}[2 \tau^2 + 3\tau +3(2\tau-1) f(\tau)],\quad
	A_{1/2} (\tau) =   2 \tau^{-2}[\tau + (\tau-1) f(\tau)],
\end{equation}
with the function $f(\tau)$ given by
\begin{equation}
	f(\tau) = \left\{
	 \begin{array}{ll}
		 \arcsin^2\sqrt{\tau},  & \quad \tau \le 1; \\
		 -\dfrac{1}{4} \left[\log\dfrac{1+\sqrt{1-\tau^{-1}}}{1-\sqrt{1-\tau^{-1}}} - i \pi\right]^2,  & \quad\tau > 1.
	\end{array}
	\right.
\end{equation}
The definitions of the dimensionless parameters are
\begin{equation}
    \tau_W = \frac{m_h^2}{ 4 m_W^2},\quad
	\tau_f = \frac{m_h^2}{4 m_f^2},\quad
    \tau_{\chii} = \frac{m_h^2}{4 m_{\chii}^2}.
\end{equation}

Based on these formulas, we can calculate the deviation of $\Gamma (h \to \gamma \gamma)$ from the SM prediction.
Green regions in Fig.~\ref{fig:23width} are expected to be excluded at 95\% CL through the $h \to \gamma \gamma$ measurement at the CEPC.
In contrast to $h$ and $Z$ invisible decays, the effect on $h \to \gamma \gamma$ via loops would not be bounded by mass thresholds.
As a result, the expected exclusion covers a large portion of the parameter space where the Higgs boson invisible decay measurement is unable to probe.

\subsection{Current experimental constraints}

\begin{figure}[!tbp]
\centering
\subfigure[~$y_1 = 0.5$, $y_2 = 1.5$. \label{fig:23cons:a}]
{\includegraphics[width=.45\textwidth,trim={0 15 0 10},clip]{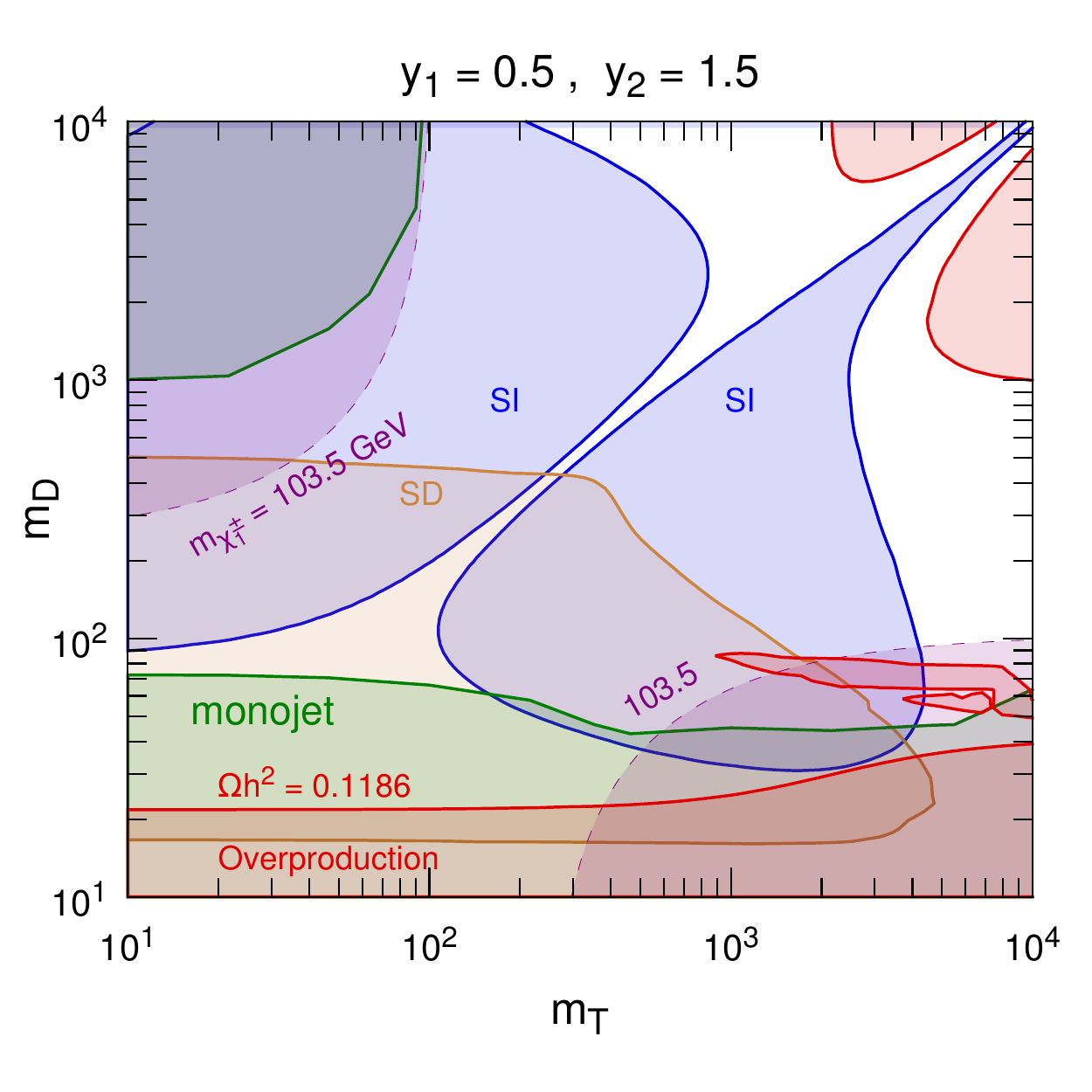}}
\subfigure[~$y_1 = y_2 = 1.0$. \label{fig:23cons:b}]
{\includegraphics[width=.45\textwidth,trim={0 15 0 10},clip]{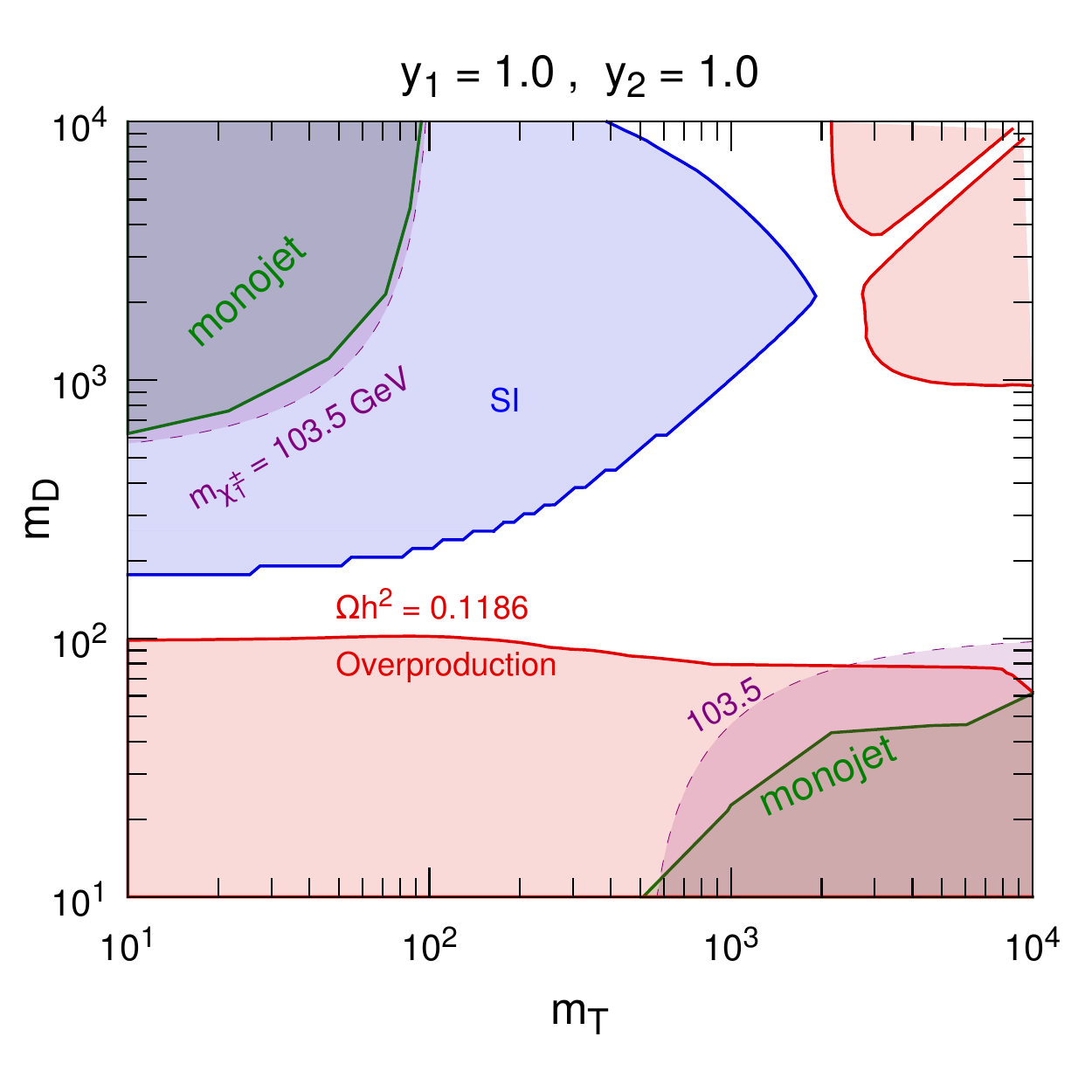}}
\subfigure[~$m_T = 100~\GeV$, $m_D = 400~\GeV$. \label{fig:23cons:c}]
{\includegraphics[width=.45\textwidth,trim={0 15 0 10},clip]{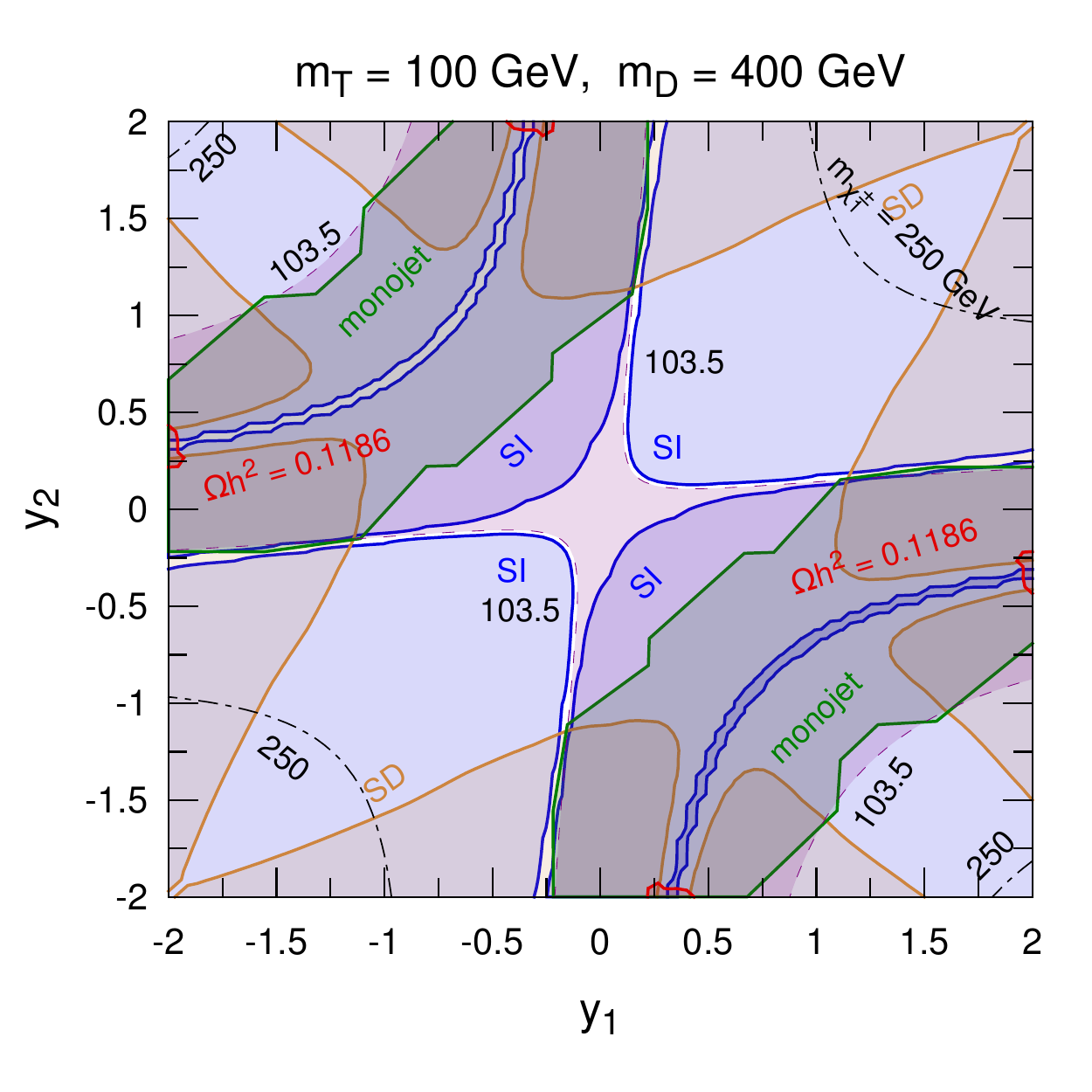}}
\subfigure[~$m_T = 400~\GeV$, $m_D = 150~\GeV$. \label{fig:23cons:d}]
{\includegraphics[width=.45\textwidth,trim={0 15 0 10},clip]{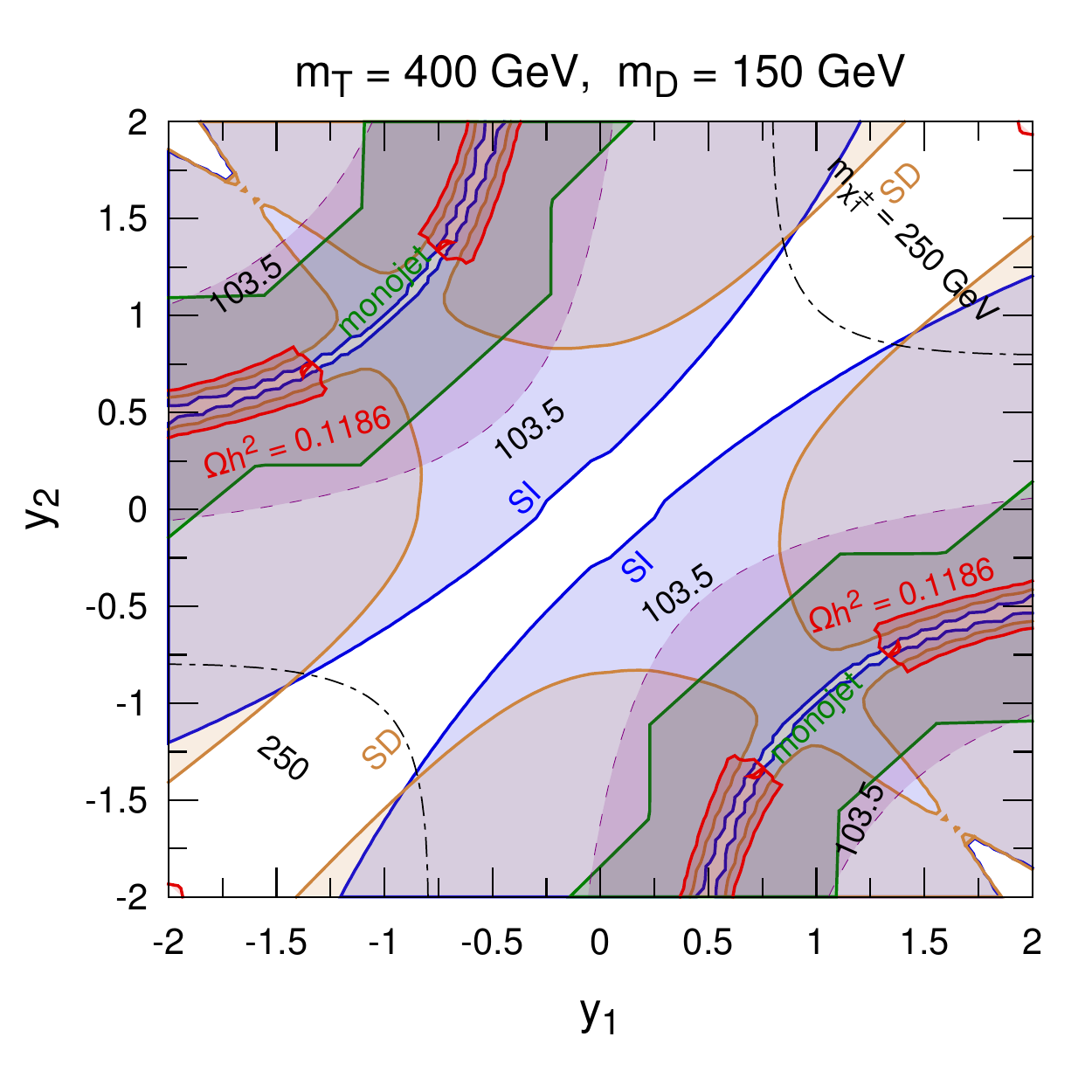}}
\caption{Experimental constraints in the $m_T-m_D$ plane (a,b) and $y_1-y_2$ plane (c,d) for the DTFDM model.
The colored regions have the same meanings as in Fig.~\ref{fig:12cons:a}.}
\label{fig:23cons}
\end{figure}

In the subsection, we discuss current experimental constraints on the DTFDM model from relic abundance, direct detection experiments, and LHC and LEP searches. Based on the study on the SDFDM model in the previous section, these calculations are quite straightforward; the results are presented in Fig.~\ref{fig:23cons}.

Red regions in Fig.~\ref{fig:23cons} indicate where DM would be overproduced in the early Universe.
Compared to the SDFDM case in Figs.~\ref{fig:12cons:a} and \ref{fig:12cons:b}, the overproduction regions with $m_D\gtrsim 1~\si{TeV}$ shrink into the corners with $m_T\gtrsim 2~\si{TeV}$ in Figs.~\ref{fig:23cons:a} and \ref{fig:23cons:b}.
This is reasonable, because the observation of relic abundance favors a DM particle mass of $\sim 2.5~\si{TeV}$ for a DM candidate purely from a fermionic triplet~\cite{Cirelli:2005uq}.
Thus, a doublet-dominated $\chia$ could saturate the universe when $m_D \gtrsim 1~\TeV$, while a triplet-dominated DM could do the same thing when $m_T \gtrsim 2~\TeV$.
This phenomenon has also been observed in the Higgsino-wino scenario of supersymmetric models~\cite{ArkaniHamed:2006mb}.
Exception occurs when $m_T \sim m_D$, where the masses the dark sector fermions are too close, leading to significant coannihilation effects that result in a much lower relic density.

Another obvious difference to the SDFDM model is that there is an overproduction regions with $m_D\lesssim 100~\si{GeV}$ for $y_1=y_2=1$ shown in Fig.~\ref{fig:23cons:b}. Unlike the SDFDM case, there is no mass degeneracy between $\chia$ and $\chiapm$ in this region, and hence the coannihilation effect is ineffective.
On the other hand, the overproduction regions in Figs.~\ref{fig:23cons:c} and \ref{fig:23cons:d} are quite small.

In Fig.~\ref{fig:23cons}, we also show the regions excluded by direct detection experiments.
The neutral mass matrices $\mathcal{M}_\mathrm{N}$ in the DTFDM and SDFDM models are identical if one treats $m_T$ and $m_S$ as the same thing.
Therefore, the neutral fermions have the same mixing pattern in the two models, which leads to identical behaviors of the $h\chia\chia$ and $Z\chia\chia$ couplings. For this reason, the SI and SD exclusion regions in Fig.~\ref{fig:23cons} have no essential difference from those in Fig.~\ref{fig:12cons}.

The exclusion limits from the ATLAS $\text{monojet}+\missET$ search are denoted by green regions in Fig.~\ref{fig:23cons}.
Electroweak production processes of two dark sector fermions in the DTFDM model are similar to \ref{eq:LHC:prod}, but now there are two charged fermions, $\chiapm$ and $\chibpm$.
The monojet search could exclude the parameter space up to $m_{\chia}\sim 80~\si{GeV}$ in Fig.~\ref{fig:23cons:a}.
In the case of $y_1 = y_2 = 1.0$ with $m_T>m_D$, however, the $Z\chi_1^0\chi_1^0$ and $h\chi_1^0\chi_1^0$ couplings vanish and there is no $pp\to \chia\chia+\text{jets}$ production.
As a result, the profile of the corresponding exclusion region in Fig.~\ref{fig:23cons:b} basically follows the contours of $m_{\chiapm}$ and $m_{\chib}$.
On the other hand, the exclusion regions in Figs.~\ref{fig:23cons:c} and \ref{fig:23cons:d} are larger than their analogues in the SDFDM model.

Pink regions in Fig.~\ref{fig:23cons} show the constraint from the LEP  searches for charged particles.
In contrast to the SDFDM model, the masses of charged fermions in the DTFDM model do not solely depend on $m_D$, but are related to all the four parameters. The exclusion regions exhibit this dependence.

\section{Conclusions and discussions}
\label{sec:conclu}

In this work, we investigate how fermionic DM affects Higgs precision measurements at the future collider project CEPC, which include the measurements of the $e^+e^-\to Zh$ cross section as well as Higgs boson invisible and diphoton decays.
In order to have influence on $e^+e^-\to Zh$ through at one-loop level, the DM particle should couple to both the Higgs and $Z$ bosons.
For this purpose, we consider two UV-complete models, SDFDM and DTFDM, where the SM is extended with a dark sector consisting of $\su2l$ fermionic multiplets. The lightest electrically neutral mass eigenstate of the additional multiplets serves as a DM candidate. Such multiplets naturally couple to electroweak gauge bosons, and their interactions with the Higgs boson come from Yukawa couplings, fulfilling our requirement.

We calculate one-loop corrections to the $e^+e^- \to Zh$ cross section induced by the dark sector in the two models.
The DTFDM model would make a bigger difference than the SDFDM model, because of stronger electroweak gauge interactions of its dark sector multiplets.
The parameter regions that could be explored via the CEPC measurement are demonstrated.
As this is a loop effect, the reachable mass scales of the dark sector would not be simply bounded by the collision energy.
For instance, CEPC with $\sqrt{s}=240~\si{GeV}$ may still be sensitive to the DTFDM model when the DM candidate mass is $\sim 900~\GeV$.

When the DM candidate is light, the Higgs boson may decay into them, resulting in an invisible decay signal. We also explore the CEPC sensitivity from such an invisible decay. But this kind of search is certainly limited by the decay kinematics.
Furthermore, the DTFDM model could affect the Higgs boson diphoton decay through quantum loops. We find that the CEPC measurement of the diphoton decay would be sensitive to much larger parameter space, compared with the invisible decay measurement.

On the other hand, these DM models are facing stringent bounds from current searches. We investigate the constraints from the DM relic abundance and DM direct detection experiments, as well as the bounds from LHC monojet searches and LEP searches for charged particles and $Z$ boson invisible decay.
We find that current experimental constraints on the two models have excluded large portions of the parameter space.
Future LHC and direct detection searches would further enlarge the corresponding exclusion regions.
Nonetheless, the full run of the high-luminosity LHC can hardly reach up to the TeV mass scales due to the low electroweak production rates.
Moreover, the models could easily escape direct detection when the parameters satisfy certain conditions, such as $y_1 \simeq y_2$ and $m_S>m_D$ ($m_T>m_D$) for the SDFDM (DTFDM) model.
The reason is that the DM couplings to the Higgs and $Z$ bosons are very weak under such circumstances. In this case, the Higgs measurements at the CEPC would be complementary to other searches.

DM annihilation in space can induce cosmic-ray and gamma-ray signals, which could be probed in DM indirect detection experiments.
Current searches have put some important constraints on the DM annihilation cross section. Interpretations of the related data depend on multiple astrophysical uncertainties, such as uncertainties from cosmic-ray propagation processes, from $J$-factors for gamma-ray fluxes, and from substructures of DM halos. In this paper, we have not used these results to constrain the models, because such constraints are not as robust as those from particle physics experiments.

It is not hard to extend this study to other models with fermionic multiplets in different $\su2l$ representations or with scalar multiplets.
Higher dimensions of representations should lead to stronger electroweak interactions and hence larger corrections to $e^+e^-\to Zh$ and $h\to\gamma\gamma$.
This kind of models, involving a dark sector with electroweak multiplets, would also have influence on $e^+e^-\to\bar{f}f$ production~\cite{Harigaya:2015yaa}, the electroweak oblique parameters~\cite{Cai:2016sjz,Cai:2017wdu}, as well as many other $e^+e^-$ production processes, such as $e^+e^-\to W^+ W^-$, $e^+e^-\to ZZ$, and $e^+e^-\to h\gamma$ production.
Furthermore, combining several such channels may be able to get a better sensitivity to the models.

\begin{acknowledgments}
This work is supported by the National Natural Science Foundation of China
under Grant Nos.~11475189 and 11475191, by the 973 Program of China under Grant No.~2013CB837000, and by the National Key Program for Research and Development (No.~2016YFA0400200). ZHY is supported by the Australian Research Council.
\end{acknowledgments}

%

\bibliographystyle{JHEP}
\bibliography{EEHZ}
\end{document}